\documentclass{aa}

\usepackage{graphicx}
\usepackage{txfonts}
\usepackage{subfig}
\usepackage{natbib}

\begin{document} 

\title{Circular polarization signals of cloudy (exo)planets}

\author{L. Rossi \and D. M. Stam}

\institute{Faculty of Aerospace Engineering, Delft University of Technology,
           Kluyverweg 1, 2629 HS Delft, The Netherlands}

\offprints{D. M. Stam
\email{d.m.stam@tudelft.nl}}

\date{Received XX, 2018; accepted YY, 2018}

\abstract
{The circular polarization of light that planets reflect
is often neglected because it is very small compared to the linear
polarization. It could, however, provide information on a planet’s
atmosphere and surface, and on the presence of life, because 
homochiral molecules that are the building blocks of life on Earth 
are known to reflect circularly polarized light.}
{We compute $P_{\rm c}$, the degree of circular polarization, of
light that is reflected by rocky (exo)planets
to provide insight into the viability of circular 
spectropolarimetry for characterizing (exo)planetary atmospheres.}
{We compute the $P_{\rm c}$ of light that is reflected by rocky (exo)planets with 
liquid water or sulfuric acid solution clouds,
both spatially resolved across the planetary disk
and, for planets with patchy clouds, 
integrated across the planetary disk, for various planetary phase angles $\alpha$.}
{The optical thickness and vertical distribution of the atmospheric gas
and clouds, the size parameter and refractive index of the cloud particles, 
and $\alpha$ all influence $P_{\rm c}$. Spatially resolved, $P_{\rm c}$ 
varies between $\pm 0.20\%$ (the sign indicates the polarization direction).
Only for small gas optical thicknesses above the clouds do
significant sign changes (related to cloud particle properties) 
across the planets' hemispheres occur. 
For patchy clouds, the disk--integrated $P_{\rm c}$ 
is typically smaller than $\pm 0.025\%$, with maximums for $\alpha$
between $40^\circ$ and $70^\circ$, and $120^\circ$ to $140^\circ$. 
As expected, the disk--integrated $P_{\rm c}$ 
is virtually zero at $\alpha=0^\circ$ and 180$^\circ$. The disk--integrated
$P_{\rm c}$ is also very small at $\alpha \approx 100^\circ$.}
{Measuring circular polarization signals appears to be challenging with 
current technology. The small atmospheric circular polarization signal
could, however, allow the detection of circular polarization due to 
homochiral molecules such as those associated with life on Earth. 
Confirmation of the detectability of such signals requires better 
knowledge of the strength of circular polarization signals of 
biological sources and in particular of the angular distribution 
of their scattering.}

\keywords{\bf Polarization --
              Techniques: polarimetric --
              Radiative transfer -- 
              Planets and satellites: atmospheres
              }

\maketitle

\section{Introduction}
\label{sec:sect_intro}

Polarimetry has been shown to be a strong tool for the characterization of 
solar system planets, with the successful characterization of the clouds of Venus
\citep{Hansen1974a}, the studies of atmospheres of other planets such as those 
of Jupiter \citep[see e.g.][and references therein]{2017A&A...601A.142M}, 
Saturn \citep{Tomasko1984}, Uranus, and Neptune \citep{Schmid2006},
and airless bodies such as comets and asteroids 
\citep[][and references therein]{kiselev2015book,Cellino2015}.
There is also a great potential for polarimetry of exoplanets because 
the light emitted by solar--type stars can be assumed to be unpolarized
when integrated across their disk \citep{Kemp1987}, 
while it is usually polarized upon reflection by a planet. Polarimetry can thus 
increase the contrast between background starlight and a planet 
\citep{Seager2000,lucas2009}, and ease the detection and confirmation 
of an exoplanet. 
And, because the degree and direction of polarization are sensitive to the
composition and structure of a planetary atmosphere and surface, polarimetry
could also help to characterize an exoplanet 
\citep[see e.g.][]{Seager2000,Karalidi2012,Stam2004,Stam2008}.

Like the state of linear polarization, the state (degree and direction) 
of circular polarization can contain information about the structure
and composition of a planetary atmosphere and about the properties of 
a planetary surface.
Circular polarization is typically induced through the multiple 
scattering of light by non-gaseous, atmospheric particles \citep{Hansen1974a}, 
or through the multiple reflection of light by a rough surface \citep{Kemp1971}, 
but, interestingly, it also displays potential for use in the detection of 
life \citep{1996P&SS...44.1441M,Bailey2000,2005AsBio...5..737S}. 
This is due to the preferred orientation of the chiral molecules
that make up organisms on Earth, referred to as homochirality 
\citep[][]{1991OLEB...21...59B}, which induces distinctive
circular polarization variations with wavelengths in reflected and transmitted
light \citep[][]
{2013EP&S...65.1167N,2012P&SS...72..111S,2009PNAS..106.7816S,2017JQSRT.189..303P}, 
although measurements on two types of cyanobacteria species by 
\citet{2016JQSRT.170..131M} also suggest that measured circular 
polarization signals can be due to internal reflections and thus not
necessarily due to homochirality. 

While many studies have concentrated on the state of the linear polarization
of light, few have considered circular polarization of reflected 
sunlight. 
Indeed, only a handful of measurements of the state of circular polarization
of solar system planets are available 
\citep{Kemp1971,1972ApJ...178..257S,1976Icar...29..235W,Kawata1978,2012P&SS...72..111S,2002OLEB...32..181M}, 
all obtained with Earth-based telescopes. As far as we know, there has not 
been a space--based instrument with circular polarization measurement 
capabilities targeting planets.
Measuring the circular polarization requires more complicated instrumentation
(often with more, and moving, parts) than measuring the linear polarization
\citep{Sparks2012}, also because the degree 
of circular polarization is usually very small, especially when 
compared to the degree of linear polarization. 
The instrument 
requirements for circular polarization instruments, in particular
the limits on the cross--talk, are thus very high. 
Recent advances in the design and production of optical elements, 
however, are increasing the viability of circular polarimetry as a 
tool for planetary characterization. In addition, the possibility of 
detecting circular polarization of light reflected by  
chiral molecules and thus possibly of detecting life, has increased the 
interest in circular polarization signals of (exo)planets.

Therefore, this study acts as a starting point for an investigation into the
feasibility of the use of circular polarimetry for planetary observations and
the identification of homochirality. 
Circular polarization can be induced by a
planetary atmosphere, surface, or a biological presence and it is important to
understand the influence that these processes have on the total circular
polarization profile of the planet. However, the degree of circular
polarization coming from the surface of the planet is irrelevant if it is
not visible due to an overwhelming signal from the atmosphere. 
Therefore, it is vital that the influence of
the atmospheric properties on the circular polarization signal 
of an exoplanet is first studied. 

In this paper we investigate the effects of clouds on the degree and
direction of circular polarization
of light reflected by Earth--like planets for both spatially resolved and
disk-integrated measurements. The latter would be representative for 
(future) measurements of starlight reflected by Earth--like exoplanets.  
In Sect.~\ref{sect2}, we describe the numerical algorithm to compute the
polarization signals and provide the physical properties of the 
atmospheres, including the clouds, and the surfaces
that cover our model planets.
In Sect.~\ref{sect3}, we present calculated polarization signals for
planets with different cloud optical thicknesses, microphysical cloud
properties, cloud top altitude, and fractional cloud coverage. 
In Sect.~\ref{sec:discussion} we discuss the limits of our simulations and the
possible issues that observers might face.
Finally, in Sect.~\ref{sec:summary}, we summarize our results.

\section{Numerical algorithms}
\label{sect2}

\subsection{Defining fluxes and polarization}

We describe the starlight that is incident on a planet and the starlight that
is reflected by the planet by a Stokes vector ${\bf F}$, as
\citep[see e.g.][]{Hansen1974,Hovenier2004}
\begin{equation}
   \mathbf{F} = \left[ \begin{array}{c}
                         F \\ Q \\ U \\ V
                         \end{array} \right],
\label{eq:def-stokes-vector}
\end{equation}
where $F$ describes the total flux, $Q$ and $U$ the linearly polarized fluxes, and
$V$ the circularly polarized flux. We express these fluxes in W m$^{-2}$. Linearly polarized fluxes $Q$ and $U$ are defined with respect to a reference
plane, for which we use the planetary scattering plane, that is, the plane that 
contains (the centre of) the star, the planet, and the observer. 
Parameters $Q$ and $U$ can straightforwardly be redefined with respect to 
another reference plane, such as the optical plane of an instrument, 
using a so--called rotation matrix 
\citep[see][for the definition]{Hovenier1983}.

The circularly polarized flux $V$ is 'the excess of flux transmitted by an 
instrument that passes right--handed circular polarization, over that transmitted 
by an 
instrument that passes left--handed circular polarization' \citep[][]{Hansen1974}. 
Regarding starlight that is reflected by a planet and that arrives at an observer, 
parameter $V$ will thus be positive if the observer 'sees' the electric vector 
rotating in the anticlockwise direction, and $V$ will be negative if the 
observer 'sees' a rotation in the clockwise direction.

We assume that the starlight that is incident on a planet 
is unidirectional and unpolarized. The justification of the latter is based on
the very small values of the disk--integrated polarization of solar-type stars. 
\citet{Kemp1987} give measurements of the linear and circular polarization of the Sun: their disk--integrated measurements yield maximal values of $0.8 \times 10^{-6}$ for the linear polarization, and $-1 \times 10^{-6}$ for the circular polarization, with a B filter. 
Locally, disturbances on the stellar disk such as stellar spots or transiting
planets could produce a break in the symmetry of the stellar disk and generate non--zero
disk--integrated values of polarization. The values expected from such
disturbances are on the order of $10^{-6}$ for the degree of linear
polarization \citep{Berdyugina2011, Kostogryz2015}.
The incident light is thus described
as ${\bf F}_0 = F_0 \bf{1}$, with $\pi F_0$ the stellar flux measured
perpendicular to the direction of propagation, and ${\bf 1}$ the
unit column vector. 

Starlight that is reflected by an orbiting planet will usually be polarized
because it has been scattered by gases and aerosols or cloud particles in the
planetary atmosphere and/or has been reflected by the surface. 
The degree of polarization of the reflected starlight is defined as 
\begin{equation}
    P = \frac{\sqrt{Q^2 + U^2 + V^2}}{F},
\label{eq:def-pol-tot}
\end{equation}
the degree of linear polarization as 
\begin{equation}
    P_{\rm l} = \frac{\sqrt{Q^2 + U^2}}{F},
\label{eq:def-pol-lin}
\end{equation}
and the degree of circular polarization as
\begin{equation}
    P_{\rm c} = \frac{V}{F}
\label{eq:def-pol-circ}
.\end{equation}
Following the description above, if $P_{\rm c} > 0$, the observer 'sees' 
the electric vector of the light rotating in the anticlockwise direction, 
and if $P_{\rm c} < 0$,
the observer 'sees' a rotation in the clockwise direction. 

\subsection{The radiative transfer algorithm}

In this paper, we will present computed (polarized) fluxes and polarization for 
both \emph{spatially resolved} and \emph{spatially unresolved}
planets. 
The latter, for which we integrate the spatially resolved signals across the 
planetary disks, apply to (future) observations of exoplanets, while the former 
apply to observations of solar system planets and also helps us to understand the 
computed exoplanet signals.

For our spatially resolved computations, we first divide the two--dimensional (2D)
planetary disk facing the observer into equal--sized, square pixels, 
with 30 or 100~pixels along the equator of the disk, which is assumed to run 
horizontally through the middle of the disk.
Then, we project the centre of each pixel onto the three--dimensional (3D) planet and 
determine the following angles for each projected pixel centre on the planet:
$\theta_0$, the angle between the local vertical and the direction towards the
illuminating sun or star, $\theta$, the angle between the local vertical and the 
direction towards the observer, and the azimuthal difference angle $\phi - \phi_0$,
which is the angle between the plane containing the local vertical and 
the propagation direction of the incident light and the plane containing the
local vertical and the direction towards the observer 
\citep[for a detailed definition, see][]{deHaan1987}.
The angles depend on the latitude and longitude of the projected pixel centre,
and $\theta_0$ and $\phi_0$ also depend on the planet's phase angle $\alpha$,
that is,\ the angle between the star and the observer as measured from the centre
of the planet ($0^\circ \leq \alpha \leq 180^\circ$).

Next, we calculate for each pixel, the Stokes vector 
(see Eq.~\ref{eq:def-stokes-vector}) of the starlight 
that is reflected at the projected pixel centre on the planet, using
\begin{equation}
   {\bf F}(\theta,\theta_0,\phi-\phi_0)= \cos \theta_0 \hspace{0.1cm}
           {\bf R}_1(\theta,\theta_0,\phi-\phi_0) \hspace{0.2cm}
                                  {\bf F}_0,
\label{eq:reflected_vector}
\end{equation}
with ${\bf R}_1$ the first column of the $4 \times 4$ local planetary 
reflection matrix. Only the first column is relevant, since the incoming 
starlight is assumed to be unpolarized. 
The actual area on the planet that is covered by a projected
pixel varies with latitude and longitude on the planet, but because all 
pixels have the same size, their respective Stokes vectors as computed 
according to Eq.~\ref{eq:reflected_vector} 
contribute equally to the planetary signal.
With our pixels covering the planetary disk, we can straightforwardly
model horizontally inhomogeneous planets by choosing different atmosphere and/or
surface models for different pixels. The physical properties of our models 
are described in Sect.~\ref{sec:sect_clouds}. 

Given an atmosphere--surface model for a given pixel,
we use an adding--doubling algorithm that 
fully includes polarization for all orders of scattering \citep{deHaan1987} 
to compute the local planetary reflection matrix
(see Eq.~\ref{eq:reflected_vector}).
Rather than embarking on a separate radiative transfer computation for every pixel,
we choose to first compute and store the coefficients 
${\bf R}_1^m(\theta,\theta_0)$
($0 \leq m < M$, with $M$ the total number of coefficients) of the expansion 
of ${\bf R}_1(\theta,\theta_0,\phi-\phi_0)$ into a Fourier series 
\citep[see][for the description of this expansion for 
the reflection matrix]{deHaan1987} for the different atmosphere--surface 
models that occur on our model planet. Our adding--doubling
algorithm computes these coefficients at values of $\cos \theta_0$ and
$\cos \theta$ that coincide with Gaussian abscissae (the total number 
of which is user--defined), and additionally for $\theta_0= 0^\circ$ and
$\theta= 0^\circ$. Given a pixel with
its local angles $\theta_0$, $\theta$, and $\phi-\phi_0$, we can 
efficiently compute its ${\bf R}_1$ by summing up \citep[see][]{deHaan1987} 
the Fourier coefficients stored for the relevant atmosphere--surface model, 
interpolating for values of $\theta_0$ and $\theta$ when necessary.

Each pixel is independent from its neighbours; there is no 3D cross--pixel
propagation of light. This remains a good approximation considering that the
projected area of each pixel on the planet is quite large, meaning that
the cross--pixel propagation will occur on scales quite small compared to the
pixel size.
Furthermore, the case most likely to be affected is that of an inhomogeneous
cloud cover (in particular at the edge of clouds). But in our simulations of
inhomogeneous covers (see Sect.~\ref{sect:inhomogeneous}), we consider averages over 300
cloud realisations, which we expect should make the effect of horizontal
propagation of light negligible.

To compute the disk--integrated Stokes vector, we sum up the Stokes vectors
pertaining to the pixels covering the planetary disk.
A locally reflected Stokes vector as computed by our adding--doubling algorithm 
is defined with respect to the local meridian plane, which contains both the
local vertical direction and the direction towards the observer. This is
perfectly fine when our aim is to compare degrees of polarization 
between different pixels, because degrees of polarization are independent of
the reference plane. However, for every pixel we redefine the local vector 
to the common reference plane (the planetary scattering plane) using the so-called rotation matrix ${\bf L}$ and $\beta$, the angle between the local meridian plane and the 
planetary scattering plane \citep[see][]{Hovenier1983}. This is done before summing up local Stokes vectors for the disk--integration. 

We normalize disk--integrated flux vectors such that the reflected 
flux at $\alpha=0^\circ$ equals the planet's geometric
albedo \citep[see][]{Stam2004,Stam2008}.
The degrees of total, linear, and circular polarization,
$P$, $P_{\rm l}$, and $P_{\rm c}$, are relative measures 
(see Eqs.\ \ref{eq:def-pol-tot} -- \ref{eq:def-pol-circ})
and thus independent of a normalization.
The degrees of polarization computed in this paper pertain only to the planet.
The polarization of the system as a whole would be lower as the star would
add mostly unpolarized flux. We ignore sources of noise such as zodiacal or
interstellar dust, or instrumental polarization.

\subsection{The model atmospheres and surfaces}
\label{sec:sect_clouds}

The atmosphere and surface of our model planets are assumed
to be locally horizontally homogeneous and flat.
Unless stated otherwise, 
the surface below the atmosphere is Lambertian, that is,\ an isotropic and 
completely depolarizing surface. Its albedo is indicated by $A_{\rm s}$. 
In this paper, we use $A_{\rm s}=0.0$ for the whole planetary surface,
unless specified otherwise. 

\begin{figure}[t]
\begin{center}
\includegraphics[width=0.8\linewidth]{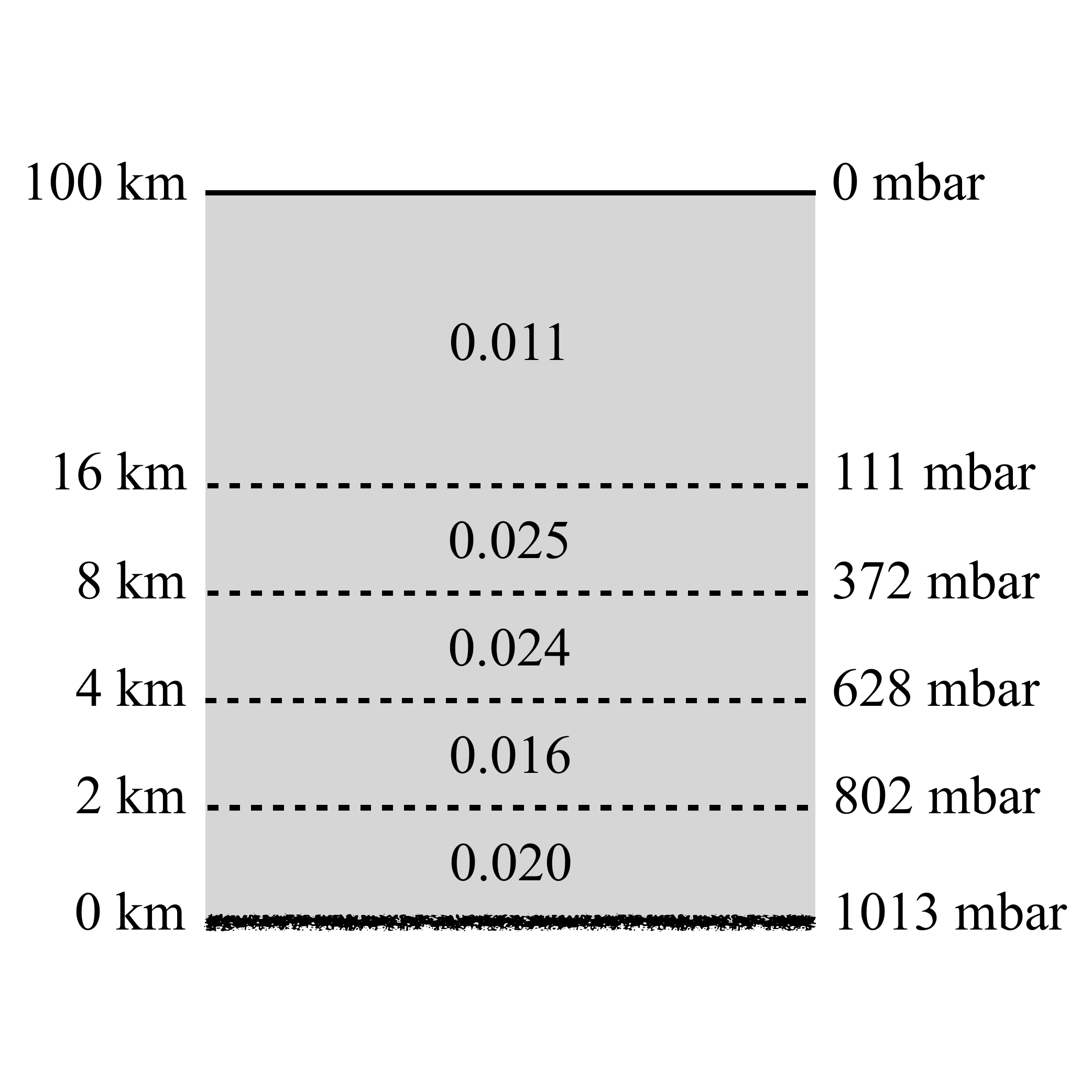}
\end{center}
\caption{Gaseous optical thickness $b_{\rm m}$ at 550~nm 
         in layers comprising the model atmosphere.
         The altitude and pressure of the levels bounding the layers are
         also given. The total gaseous optical thickness of the model
         atmosphere is 0.096.}
\label{fig:atmos}
\end{figure}

The atmosphere can be vertically inhomogeneous: it is composed of a stack of
homogeneous layers. Each layer contains gaseous molecules that are anisotropic
Rayleigh scatterers with a depolarization factor of 0.0279
\citep[see][]{Hansen1974}, and, optionally, cloud particles. The amount of gas
and cloud particles in a layer is represented by the optical thicknesses
$b_{\rm m}$ and $b_{\rm c}$, respectively. The total gaseous optical thickness
of our model atmosphere (thus summed over all layers) is 0.096, which is
representative for the Earth's atmosphere at a wavelength of about 550~nm
\citep[see][for a detailed description of how $b_{\rm m}$ is
computed]{Stam2008}.  We ignore absorption by atmospheric gases.
Figure~\ref{fig:atmos}
shows how this optical thickness is divided over five atmospheric layers.  The
cloud particles are distributed homogeneously within one of these layers.
 
The single scattering matrix of most types of atmospheric scatterers is 
given by \citep[see][]{Hansen1974}
\begin{equation}
{\bf P}(\Theta) = \left[ \begin{array}{cccc}
                    P_{11}(\Theta) & P_{21}(\Theta) & 0 & 0 \\
                    P_{21}(\Theta) & P_{22}(\Theta) & 0 & 0 \\
                    0 & 0 & P_{33}(\Theta) & -P_{34}(\Theta) \\
                    0 & 0 & P_{34}(\Theta) &  \hspace{0.2cm} P_{44}(\Theta) 
                  \end{array}  \right],
\label{eq:scatmat}
\end{equation}
with $\Theta$ the single scattering angle ($0^\circ\leq\Theta \leq 180^\circ$,
with $\Theta=0^\circ$ indicating forward scattered light).

For (anisotropic) Rayleigh scattering, matrix element $P_{34}$ equals zero. 
Consequently, if Rayleigh scattering is the only scattering process in a 
planetary atmosphere, incident unpolarized light cannot become circularly
polarized by scattering in the atmosphere, not even when it is scattered
multiple times. For atmospheric particles that are large compared
to the wavelength of the incident light, such as cloud particles, 
element $P_{34}$ is non--zero for most
values of $\Theta$. Unpolarized incident light can thus become circularly
polarized when it is scattered at least twice by such particles (or 
first at least once by a Rayleigh scatterer and then by a large aerosol
or cloud particle).

We assume that the particles constituting the clouds are spherical and
internally homogeneous. We compute their single scattering matrix with the 
Mie--scattering algorithm described by \citet{deRooij1984}.
The sizes of the cloud particles follow the two-parameter gamma 
distribution, with an effective
radius $r_{\rm eff}$ and an effective variance $v_{\rm eff}$. 
Our standard values for $r_{\rm eff}$ and $v_{\rm eff}$ are 8.0~$\mu$m
and 0.1, respectively, similar to Earth cloud values from \citet{Han1994}. 
For the refractive index we use $1.33+10^{-8}i$, which is 
representative for liquid water at 550~nm \citep{Hale1973}.
The single scattering albedo of these particles is close to one.
We will also use Venus--like cloud particles, described by the same 
size distribution, except with $r_{\rm eff}= 1.05~\mu$m, $v_{\rm eff}=0.07$,
and with a refractive index of $1.44+0.015i$, which is representative
for a sulfuric acid solution at 500~nm \citep{Hansen1974a}.
The single scattering albedo of the Venus--like particles is 0.75.

\begin{figure}[t]
\begin{center}
\includegraphics[width=9cm]{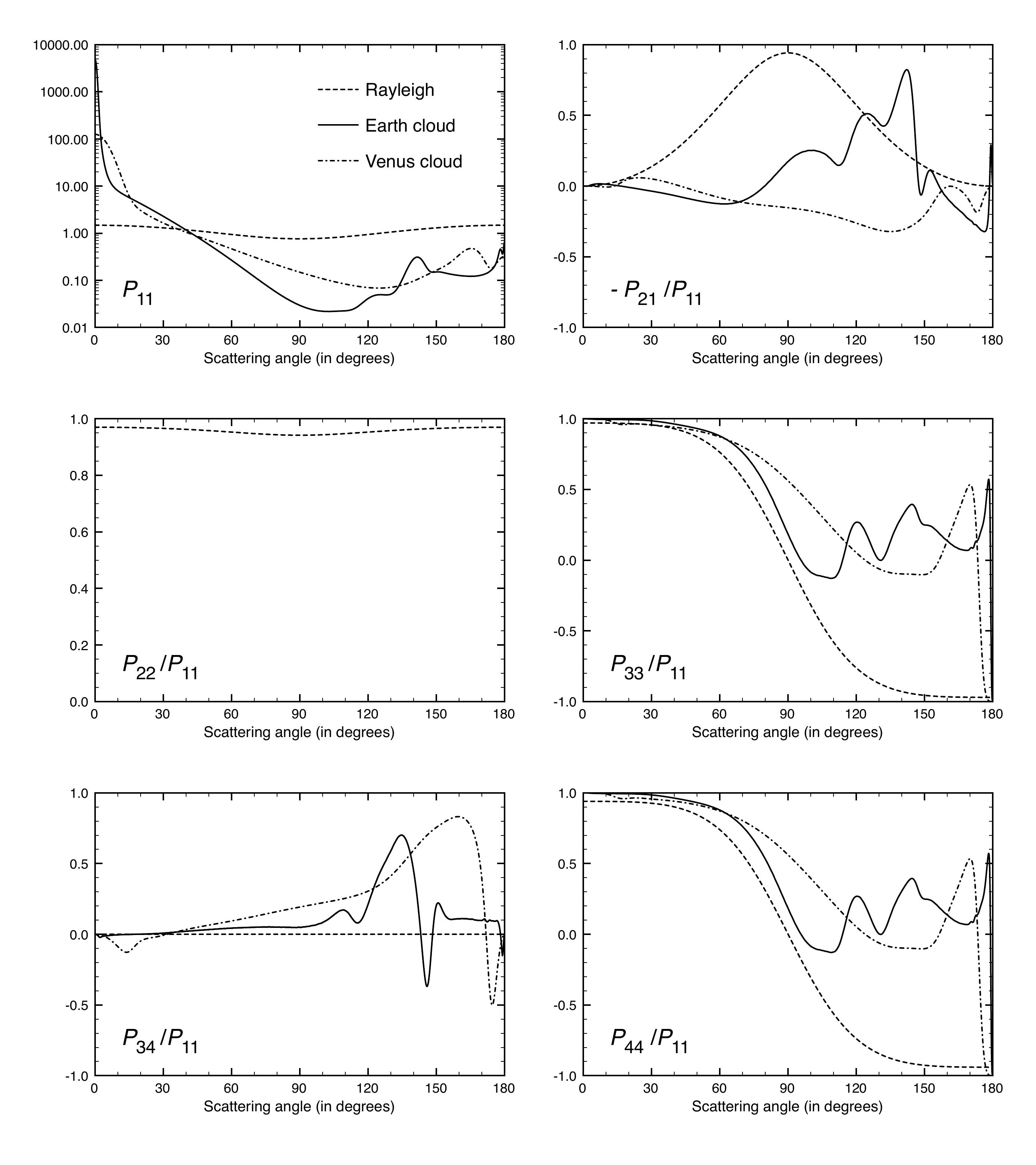}
\end{center}
\caption{Single scattering matrix elements of Earth--like
         cloud particles (solid lines), Venus--like cloud particles 
         (dashed-dotted lines),
         and Rayleigh scattering molecules (dashed lines) at 0.55~$\mu$m. 
         Elements $P_{11}$ have been normalized such that their averages 
         over all scattering angles equal one, and the other elements
         have been normalized to $P_{11}$.
         For spherical particles, such as the cloud 
         particles, $P_{22}/P_{11}=1$, $P_{34}= -P_{43}$, and 
         $P_{33}=P_{44}$.
         The Earth--like cloud particles are described by a two-parameter 
         gamma size distribution with $r_{\rm eff}=8.0$~$\mu$m and
         $v_{\rm eff}=0.1$. Their refractive index is 
         $1.33+10^{-8}i$ \citep{Hale1973}. The Venus--like cloud
         particles have $r_{\rm eff}= 1.05~\mu$m, $v_{\rm eff}=0.07$,
         and a refractive index of $1.44+0.015i$ \citep{Hansen1974a}.}
\label{fig:single}
\end{figure}

Figure~\ref{fig:single} shows the elements of the scattering matrix
for the two types of cloud particles. As can be seen from 
Eqs.~\ref{eq:def-pol-lin} and \ref{eq:scatmat},
the ratio $P_{21}/P_{11}$ describes the degree 
of linear polarization of incident unpolarized light that has been scattered 
only once.
In Fig.~\ref{fig:single}, we have included a minus sign for this ratio
to indicate the direction of polarization of the scattered light:
if $-P_{21}/P_{11} > 0$, the scattered light is polarized perpendicular to
the plane that contains both the incoming and the scattered beams, 
while if $-P_{21}/P_{11} < 0$, it is polarized parallel to this plane.

The curves in Fig.~\ref{fig:single} clearly show the strong forward 
scattering peak in the phase functions (elements $P_{\rm 11}$) of both 
cloud particle types, and the enhanced scattering
representative for the first and second order rainbows, around 
$\Theta= 140^\circ$ and 125$^\circ$, respectively, in the phase
function of the water cloud particles. These rainbow 
features also show up in other matrix elements, in particular 
in $-P_{21}/P_{11}$.

\subsection{The cloud cover}

\begin{figure*}
\centering
\includegraphics[width=14cm]{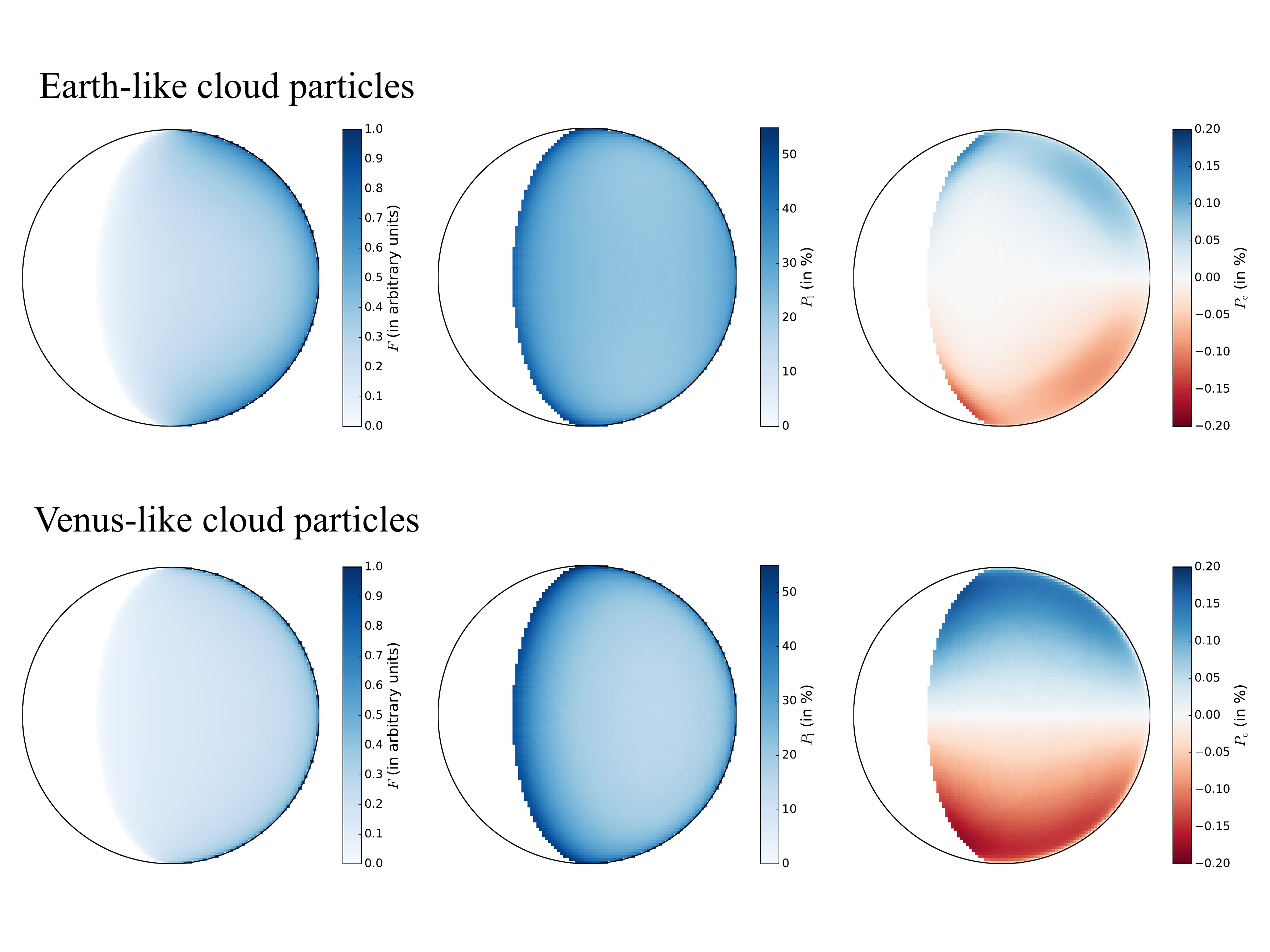}
\caption{Reflected light signals of planets with horizontally homogeneous 
         clouds composed of Earth--like cloud particles (top row) or Venus--like 
         cloud particles (bottom row),
         at a phase angle $\alpha$ of 60$^\circ$. For each model planet,
         the cloud optical
         thickness $b_{\rm c}$ is 2.0, the cloud bottom is at 2.0~km, its
         top at 4.0~km (see Fig.~\ref{fig:atmos}), and the surface is black.
         Left column: The total flux $F$ (normalized to the maximum on the disk);
         middle column: The degree of linear polarization $P_{\rm l}$;
         right column: The degree of circular polarization $P_{\rm c}$. 
         Integrated across the planetary disk, $P_{\rm l}=25.5 \%$ for the 
         Earth--like clouds and 23.8 $\%$ for the Venus--like clouds, in both
         cases with the polarization direction perpendicular to the planetary 
         scattering plane.}
\label{fig3}
\end{figure*}

To investigate the sensitivity of the circularly polarized flux due to the
cloud optical thickness, cloud particle properties (size and composition), 
and cloud top altitude,
we will show fluxes and degrees of polarization calculated for planets that 
are completely covered by a cloud deck.
The disk--integrated circularly polarized flux (and degree of circular
polarization) of a horizontally homogeneous planet will be zero, because
the circularly polarized flux of a pixel on the 
northern hemisphere will be cancelled by that of the corresponding pixel on 
the southern hemisphere, as that would have the same absolute value but the 
opposite sign (this can be seen in Fig.~\ref{fig3}). 
Thus, to simulate the circularly polarized fluxes of spatially unresolved
planets, such as exoplanets,
we will use horizontally inhomogeneous model planets that are asymmetric with 
respect to the planetary scattering plane due to patchy clouds. 
We will limit ourselves to a single type of cloud per planet, thus, the clouds
in all cloudy pixels have the same optical thickness, altitude, and 
particle properties.

The patchy clouds are described by $f_{\rm c}$, the fraction of all pixels
on the whole disk that are cloudy, and by the actual distribution of the cloudy
pixels across the disk.
The patches are generated with the method described by \citet{Rossi2017}, 
where cloudy pixels are defined by a 2D Gaussian, randomly placed on the 
square grid and which shape and 
orientation are chosen to simulate streaky, zonally--oriented, clouds.
Figure~\ref{fig:patches} shows examples of cloud patterns for  
$f_{\rm c}=0.3$ (i.e.\ 30~\%~cloud cover) and $f_{\rm c}=0.5$.

For each cloud cover pattern across a planet,
we define the asymmetry factor $\gamma$,
which indicates the percentage of pixels (cloudy and non--cloudy) 
on one of the illuminated and visible hemispheres that \emph{do not have} 
a similar pixel in a mirrored position on the other hemisphere 
(assuming that the planetary scattering plane divides the two
hemispheres). If $\gamma=0.0$, all pixels have a mirror pixel
(the circular polarization signal of this planet will equal zero),
while if $\gamma=1.0$, none of the pixels has a mirror pixel.
The asymmetry factors for the examples shown in Fig.~\ref{fig:patches}
are 0.24 and 0.52 for the two $f_{\rm c}=0.3$ cases, and 0.28 and 0.61
for the two $f_{\rm c}=0.5$ cases.


\section{Results}
\label{sect3}

In this section, we will first show and discuss the reflected total and 
polarized fluxes for horizontally homogeneous planets
(Sect.~\ref{sect:homogeneous}), and then we will investigate the effects
of horizontal inhomogeneities due to fractional cloud coverage
(Sect.~\ref{sect:inhomogeneous}). For horizontally homogeneous planets,
$P_{\rm c}$ will be zero when integrated across the planetary disk. For
these planets, we will therefore only show and discuss signals of spatially 
resolved planets. Horizontally inhomogeneous planets can have a non--zero 
disk--integrated value of $P_{\rm c}$. The spatially resolved signals 
of these planets can straightforwardly be derived from the spatially 
resolved signals of the horizontally homogeneous planets. For the horizontally 
inhomogeneous planets, we will thus only discuss the spatially unresolved signals,
which would be representative for exoplanet observations.

\subsection{Horizontally homogeneous planets}
\label{sect:homogeneous}

As mentioned earlier, a planet with a purely gaseous atmosphere  
and a depolarizing or linearly polarizing surface will not reflect 
any circularly polarized flux: circularly polarized flux will only appear when 
the atmosphere contains aerosol and/or cloud particles.
Figure~\ref{fig3} shows the reflected flux $F$, and the degrees of linear
and circular polarization, $P_{\rm l}$ and $P_{\rm c}$, respectively, for a 
planet at a phase angle $\alpha=60^\circ$ (the single scattering angle $\Theta$
is thus 120$^\circ$ across the planetary disk). 
The horizontally and vertically homogeneous cloud layer has an optical 
thickness $b_{\rm c}= 2.0$ (at 550~nm), its bottom at 2.0~km, and its 
top at 4.0~km. The clouds are composed of either the Earth--like or 
the Venus--like particles. 

It can be seen that for both particle types, 
the reflected flux $F$ is fairly constant 
across the planetary disk, thanks to the homogeneous layer of clouds that 
hides the dark surface. At the limb, $F$ (per pixel) is largest, 
because there the reflected light has been mostly scattered by the gas 
above the cloud, and the scattering phase function of gas molecules is 
relatively large at this phase angle
angle (i.e.\ a scattering angle of 120$^\circ$, see Fig.~\ref{fig:single}). 
At the terminator, the incident flux per atmosphere surface area  
is very small, yielding small values of $F$.

The degree of linear polarization, 
$P_{\rm l}$, is also fairly constant across the disk, because it is strongly
determined by the singly scattered light, and the single scattering angle
is constant across the disk. Values of $P_{\rm l}$ are somewhat higher at
the limb and the terminator for both types of cloud particles
because there the contribution of multiple scattered light from the lower
layers, with usually a 
low degree of polarization, is relatively small, and there is a stronger 
contribution of relatively highly polarized light scattered by the molecules higher in the atmosphere.
The smallest local values of $P_{\rm l}$ for the Venus--like cloud particles are 
smaller than those for the Earth--like cloud particles. That is due to 
the difference in their single scattering polarization at this phase angle
(cf. Fig.~\ref{fig:single}): indeed, were it not for the gaseous molecules overlying 
the cloud layer, the Venus--like cloud particles would yield a planet with
a polarization direction parallel to the reference plane. Both the gas molecules
and the Earth--like cloud particles yield perpendicular polarization.
Integrated across the planetary disks, $P_{\rm l}=25.5~\%$
for the planet with the Earth--like clouds and $24.5~\%$ for the planet with
the Venus--like clouds at this phase angle.

The degree of circular polarization, $P_{\rm c}$, depends on the multiple
scatterings involving a cloud particle at least once.
As expected, it is generally small for both types of cloud particles,
and the pattern of $P_{\rm c}$ on the 
northern hemisphere mirrors that on the southern hemisphere, except 
with the opposite sign, because of the horizontal homogeneity of the cloud
cover. Along the equator, $P_{\rm c}$ equals zero due
to symmetry. Because of this antisymmetry of the circular polarization on the planetary disk, $P_{\rm c}=0 \%$ for
both planets when integrated across the planetary disk.

Through the multiple scattering, $P_{\rm c}$ depends on the illumination and 
viewing geometries (hence on the phase angle and the location on the planet)
and on the composition and structure of the planetary atmosphere and surface
(the latter is black for our model planets).
In particular, the gas molecules above the clouds appear to have a significant
influence on $P_{\rm c}$, because they scatter linearly polarized light
onto the clouds, and the cloud particles can
subsequently scatter light back to space in a circularly polarized state, through
the single scattering matrix element $P_{\rm 43}$ (see Fig.~\ref{fig:single}). 
Because this element has the same sign for the Earth--like and the Venus--like
cloud particles, $P_{\rm c}$ across the planet has the same sign for both cloud
types. The higher value of element $P_{\rm 43}$ of the Venus--like particles
is responsible for the higher values of $P_{\rm c}$ across the planet with
the Venus--like clouds.
Another reason for the difference in the absolute value of $P_{\rm c}$ is the 
absorption within the Venus--like cloud particles (their single scattering
albedo is 0.75 versus almost 1.0 for the Earth--like particles), 
which suppresses the multiple scattering of light between cloud particles.
Because multiple scatterings between cloud particles randomize and hence 
usually lower the degree of polarization, less multiple scatterings will 
usually yield a higher $P_{\rm c}$.

The relative importance of the scattering by the gas and the cloud layer 
on $P_{\rm c}$ can be analysed more in depth by unravelling the different
contributions.
Figure~\ref{fig4} shows $P_{\rm c}$ for five model atmospheres with varying
optical thicknesses of the gas and the cloud, for the two cloud types.
The vertical extension of the cloud is from 2.0 and 4.0~km, like in 
Fig.~\ref{fig3}. When filled with gas (models A, B, and C), the optical 
thickness of the lowest layer is 0.020, otherwise (models D and E) it is zero. 
The gaseous optical thickness of the second layer is zero (models C and D) 
or 0.016 (models A, B, and E), and that of the third layer is zero 
(models B, C, and D) or (cf.\ Fig.~\ref{fig:atmos}) 
0.024 + 0.025 + 0.011 = 0.060 (models A and E).
By comparing the signals for the different model atmospheres, 
it is clear that the gas above the
cloud strongly influences $P_{\rm c}$: the linearly polarized light
scattered downwards by the gas yields circularly polarized light through
the single scattering matrix element $P_{34}$ (which equals -$P_{43}$) of the 
cloud particles for models A and E. 
Because element $P_{34}$ has the same sign across most of the 
single scattering angle range for the two cloud particle types 
(see Fig.~\ref{fig:single}) and because the incident linearly polarization
field of the single scattered light is the same, the sign of 
$P_{\rm c}$ is the same for the two types of planets.

For the atmospheres without gas above the cloud (models B, C, and D), 
$P_{\rm c}$ shows a pattern that depends on the cloud particle type, as
it is mostly due to light that has been scattered twice by the cloud
particles. The difference in the $P_{\rm c}$ pattern is to be expected 
because the direction of linear polarization of the 
singly scattered light is different for the different cloud particle types 
across most of the single scattering angle range,
while the direction of circular polarization is the same across the scattering
angle range for the different cloud particle types (cf. Fig.~\ref{fig:single}).
In the absence of (a significant amount of) gas above the clouds, the pattern of 
$P_{\rm c}$ across the 
planet will also depend slightly on the gas within and/or below the clouds. 
This can be seen by comparing the plots for models B, C, and D in 
Fig.~\ref{fig4}.
This gas within and/or below the clouds can scatter linearly polarized 
light towards the cloud particles that can subsequently be scattered 
as circularly polarized light, and the light scattered by the gas can 
dilute $P_{\rm c}$ by adding unpolarized light to the signal emerging 
from the atmosphere.

Next, we investigate the effects of the vertical position of a cloud
in a planetary atmosphere (with gas in all atmospheric layers). This position 
influences $P_{\rm c}$
because it influences the optical path lengths in the atmosphere, and hence for example\
the amount of linearly polarized light that is incident on the cloud from 
above and/or from below.
Figure~\ref{fig5} shows the spatially resolved $P_{\rm c}$ for the two cloud types with 
$b_{\rm c}= 2.0$ with a cloud top altitude $z_{\rm t}$ equal to 
2, 4, 8, 16, and 100~km (see Fig.~\ref{fig:atmos},
the cloud fills its atmospheric layer). A cloud top altitude of 100~km,
thus at the top of the atmosphere, is unrealistic, but has been included to 
gain insight in the light scattering processes. 

Figure~\ref{fig5} shows that increasing $z_{\rm t}$ up to at least $z_{\rm t}=8$~km,
leads to a decrease of $P_{\rm c}$ for both cloud types. This decrease is due 
to the decrease of the linearly polarized, Rayleigh scattered flux that is incident 
on the top of the cloud. When $z_{\rm t}$ is increased even further,\ 
to 16 and 20~km, the contribution of the gas below the cloud layer   
to $P_{\rm c}$ becomes significant. The difference in the amount of gas 
below the cloud is responsible for the difference in the patterns between 
the planet plots for model B in Fig.~\ref{fig4} (a low cloud, hence little 
gas below the cloud) and those for $\alpha=60^\circ$
and $z_{\rm t}=100$~km in Fig.~\ref{fig5}.

Not only the gas optical thickness influences the optical path through the 
atmosphere and hence $P_{\rm c}$, but also the cloud optical thickness $b_{\rm c}$.
Figure~\ref{fig7} shows $P_{\rm c}$ 
for Earth--like and Venus--like particles, and for 
$b_{\rm c}$ equal to 0.5, 1.0, 2.0, 4.0, and 10.0
at $\alpha$ equal to 30$^\circ$, 60$^\circ$, 90$^\circ$, and
120$^\circ$. The cloud top in all cases is at 4~km.
For the model atmospheres with Earth--like cloud particles, $P_{\rm c}$ 
seems largest for $b_{\rm c}= 2.0$. For smaller $b_{\rm c}$,
the amount of multiple scattering is too small to induce significant circularly
polarized flux, while for increasing $b_{\rm c}$ above 2.0, the randomizing 
influence of the increasing multiple scattering decreases $V$ while increasing
$F$, and thus decreases $P_{\rm c}$. The latter effect seems smallest at 
$\alpha=120^\circ$, because there the optical path lengths for the
light reaching the observer are already very long and increasing $b_{\rm c}$
does not significantly change the contribution of multiple scattered light to the
received signal.

The atmospheres with the Venus--like cloud particles show a different
behaviour at least for $\alpha=30^\circ$, 60$^\circ$, and 90$^\circ$: $P_{\rm c}$
increases with increasing $b_{\rm c}$ until $b_{\rm c}=2.0$, and then remains
more or less constant. The reason is that the Venus--like cloud particles absorb 
light, and hence prevent a
strong increase in multiple scattered light with increasing $b_{\rm c}$ 
and thus prevent a decrease of $P_{\rm c}$.

\begin{figure}
\begin{center}
\includegraphics[width=8cm]{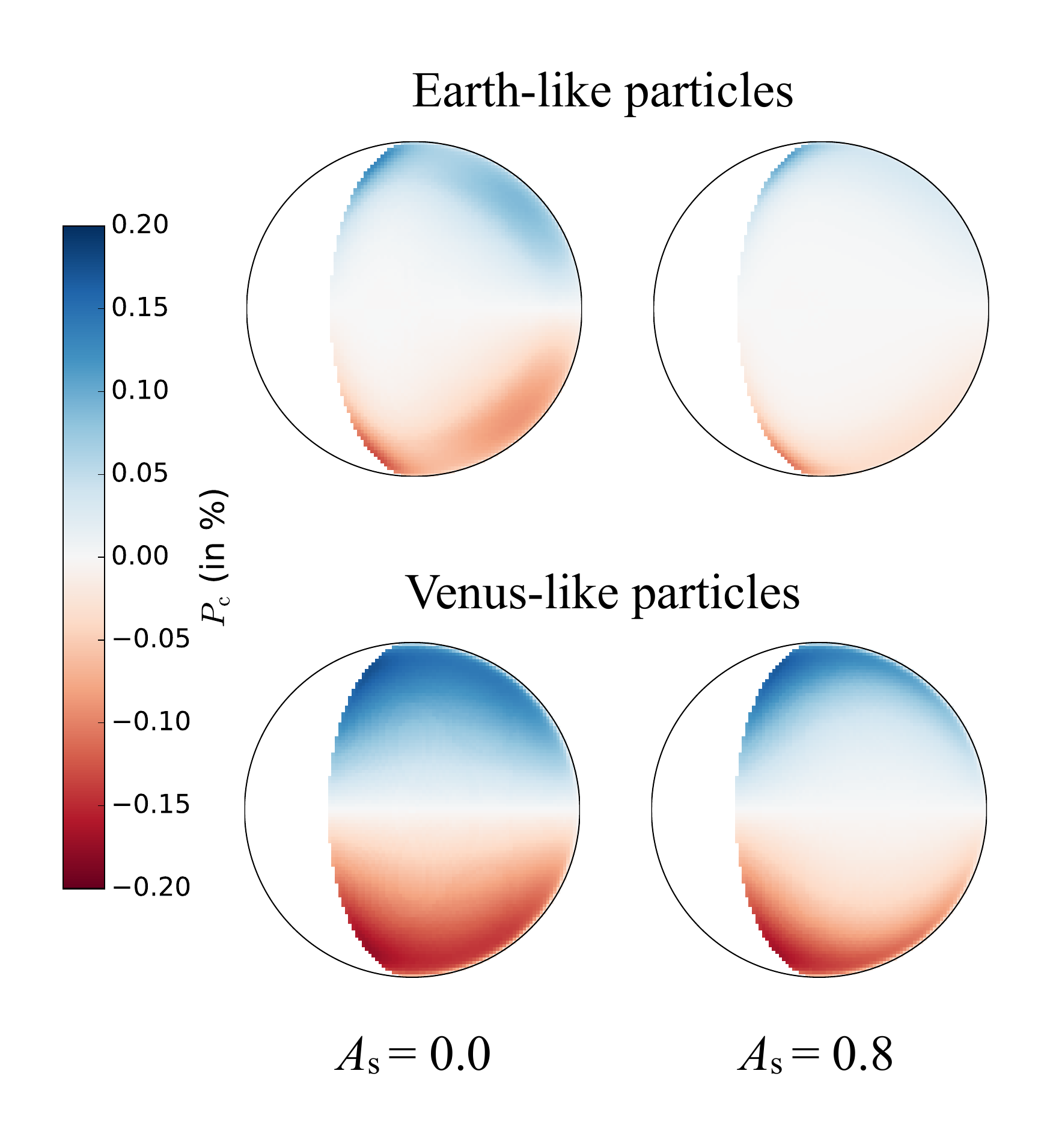}
\end{center}
\caption{Degree of circular polarization $P_{\rm c}$ for the standard 
         model atmosphere with the cloud top at 4~km, $b_{\rm c}$ equal to 
         2.0, and Earth--like (top) and Venus--like (bottom) cloud particles.
         The phase angle is 60$^\circ$, and the surface albedo $A_{\rm s}$
         equals 0.0 (left column) or 0.8 (right column).}
\label{fig8}
\end{figure}

In the results presented so far, the model atmospheres were bounded below by
black surfaces. The influence of a reflecting surface on $P_{\rm c}$ 
is shown in Fig.~\ref{fig8},
where $P_{\rm c}$ across the planet has been plotted for surface albedo's 
$A_{\rm s}$ equal to 0.0 (the standard model) to 0.8. The surface reflection 
is Lambertian, that is,\ isotropic and unpolarized. The atmosphere is the 
standard model atmosphere (see Fig.~\ref{fig3}). The cloud has its top at 
4.0~km and $b_{\rm c}= 2.0$.
As expected, increasing $A_{\rm s}$ adds unpolarized light to the
bottom of the cloud layer and decreases $P_{\rm c}$. This decrease of $P_{\rm c}$
with increasing $A_{\rm s}$ decreases with increasing $b_{\rm c}$.

In case the surface below the atmosphere reflects linearly polarized light,
it can influence the pattern of $P_{\rm c}$ across the planet, as this light
can be scattered by the cloud particles and become circularly polarized,
but the overall effect is expected to be small because in order to have 
a significant effect, there should be a horizontally homogeneously 
polarizing surface in combination with a high surface albedo. 
Even if these surface characteristics were met, a cloudy atmosphere 
would lead to a relatively low and diffuse surface illumination, limiting the
absolute amount of linearly polarized flux that would reach the clouds,
especially with larger values of $b_{\rm c}$.
A circularly polarizing surface, for example covered by vegetation or other 
organic materials \citep[e.g.][]{2017JQSRT.189..303P,2012P&SS...72..111S}, 
could directly contribute to a planet's circularly
polarized flux, thus without requiring an additional scattering by cloud
particles. However, if the surface reflection were independent of the
azimuthal angle, for example, if the angular distribution of the reflected light
were isotropic, the surface would leave no net circular polarization signal 
at the top of the atmosphere.

\subsection{Horizontally inhomogeneous planets}
\label{sect:inhomogeneous}

We next investigate the influence of patchy cloud covers and the cloud cover
variability on $P_{\rm c}$ of starlight that is reflected by a spatially 
unresolved planet, such as an exoplanet. 
As discussed above, the disk--integrated $P_{\rm c}$ of a horizontally homogeneous 
planet is zero, as contributions from one hemisphere cancel out those of the other 
hemisphere. Therefore, only horizontally inhomogeneous planets would yield 
non--zero values of $P_{\rm c}$.
As Figs.~\ref{fig5}--\ref{fig8} show, $P_{\rm c}$ can vary significantly
across a horizontally homogeneous cloudy planet, while it is zero in the absence 
of clouds. So patchy clouds could yield a net amount of circular polarization
with $P_{\rm c}$ depending on the distribution of the patches across the 
planetary disk.
Indeed, even for patchy clouds, any symmetry with respect to the scattering plane
will cancel the net circular polarization. A strongly symmetric 
cloud cover across the planet would yield a relatively 
low net circular polarization signal, while the most extreme case,\ a planet with one fully cloudy and one 
cloud--free hemisphere (asymmetry parameter $\gamma$ would thus equal 1.0),
could yield a relatively large net signal.

\begin{figure}[h!]
    \centering
    \includegraphics[width=.45\linewidth]{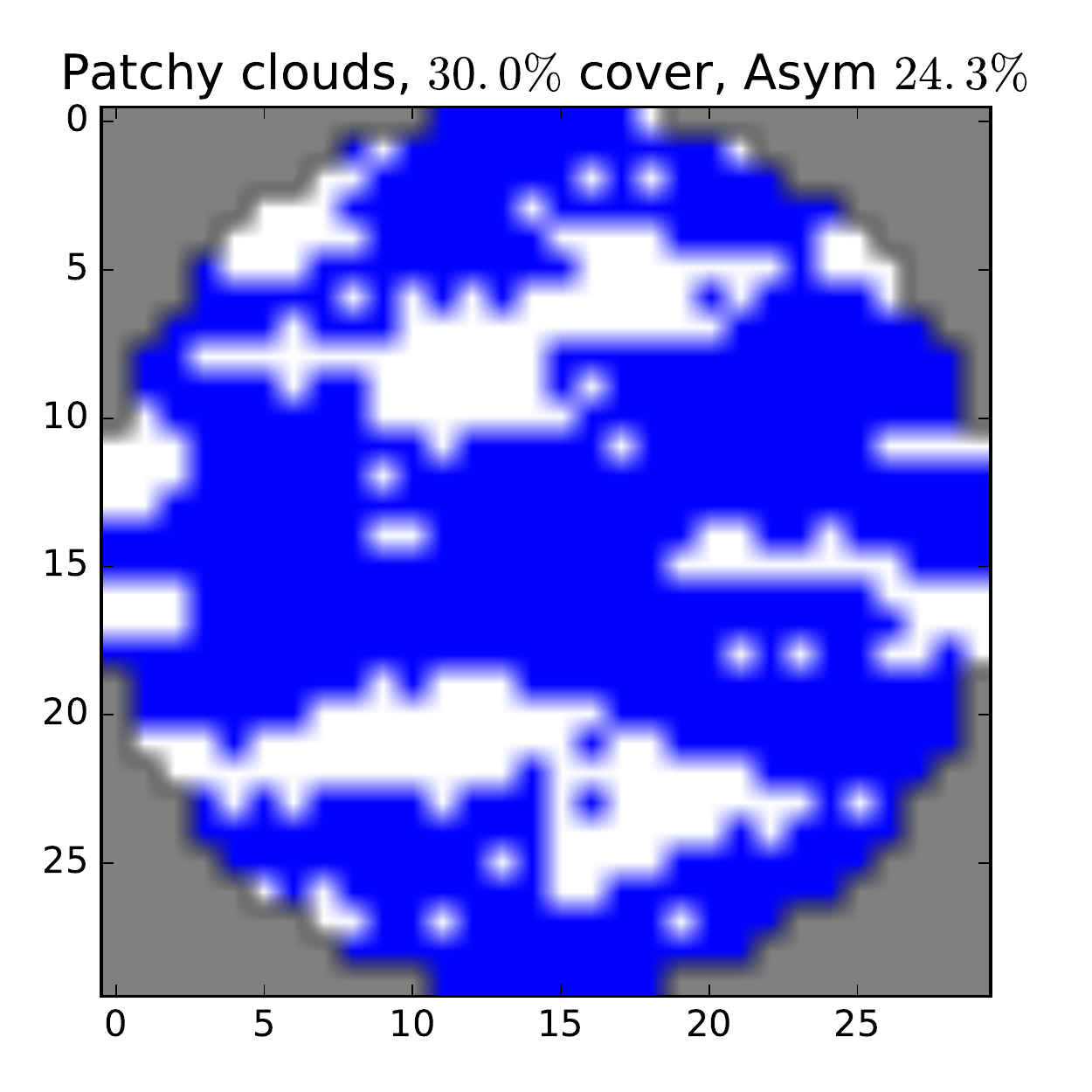}
    \includegraphics[width=.45\linewidth]{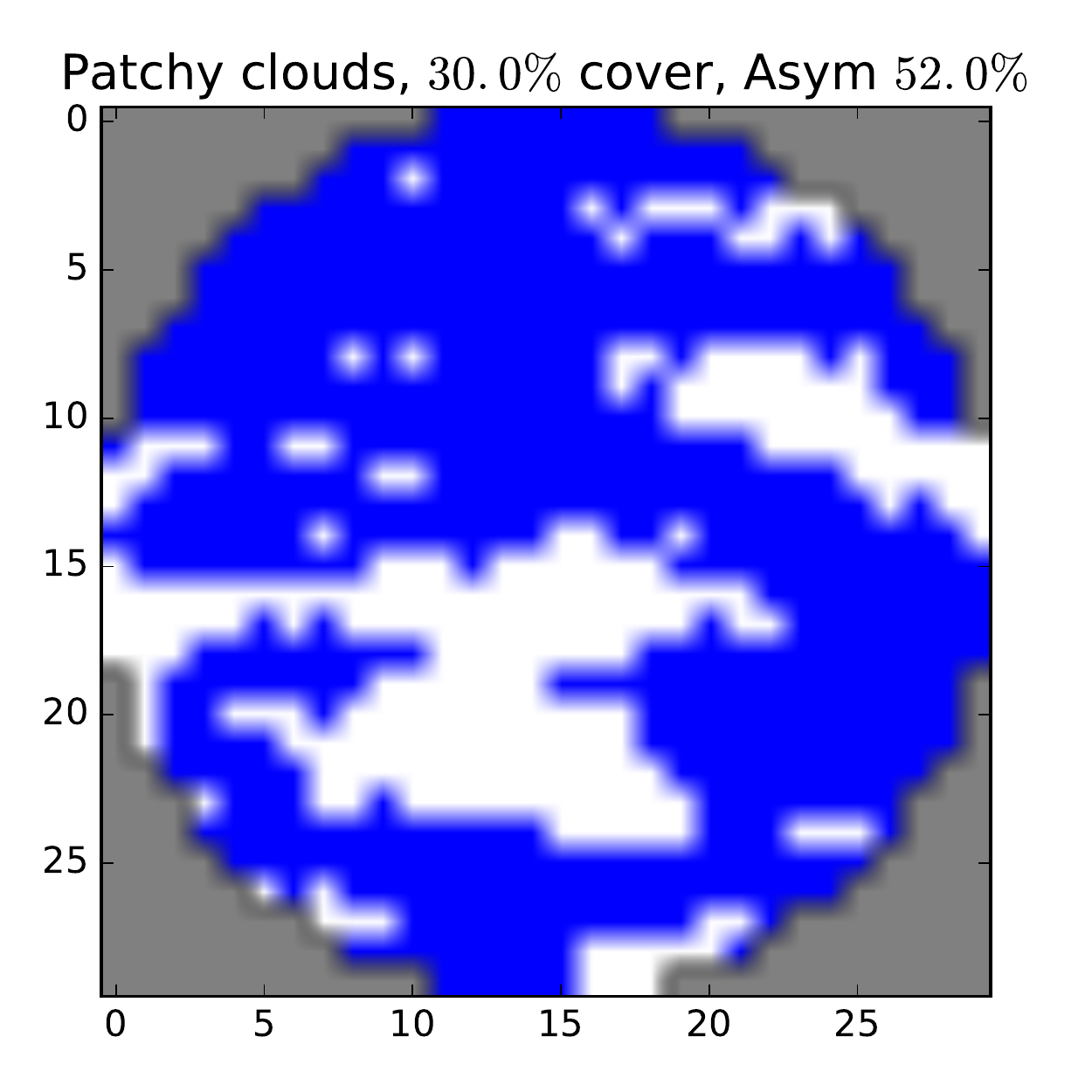}

    \includegraphics[width=.45\linewidth]{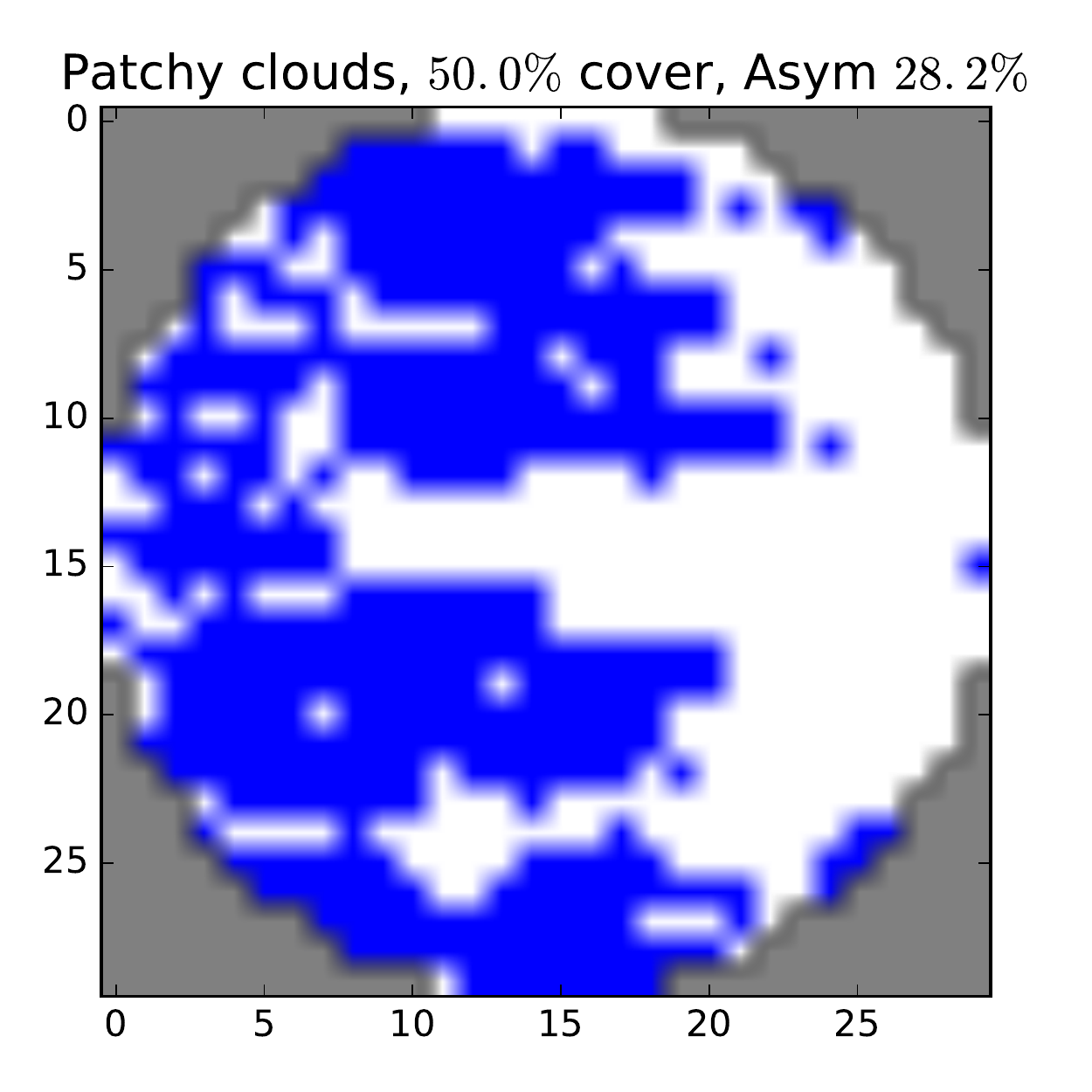}
    \includegraphics[width=.45\linewidth]{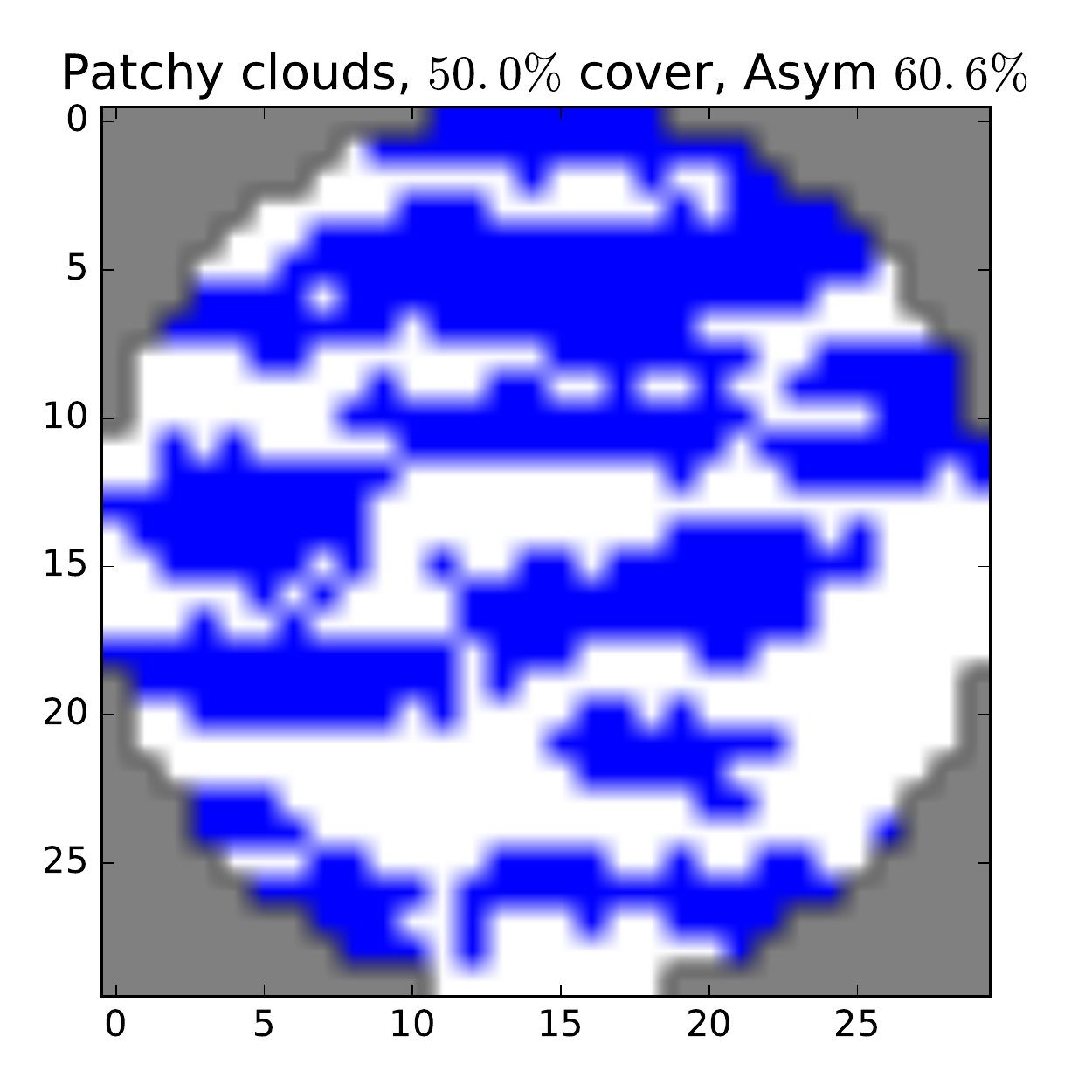}
    \caption{Examples of patchy cloud patterns for $f_{\rm c}= 0.3$ (top) 
             and $f_{\rm c}= 0.5$ (bottom) at $\alpha=0^\circ$. 
             The asymmetry factors $\gamma$ of these planets 
             are 0.24 (top--left), 0.52 (top--right), 0.28 (bottom--left) 
             and 0.61 (bottom--right).}
\label{fig:patches}
\end{figure}

We use the cloud algorithm described in Sect. \ref{sec:sect_clouds} \citep[see ][for a more detailed explanation]{Rossi2017} to generate 
300~different patchy cloud covers for various values of the cloud 
coverage fraction $f_{\rm c}$ and compute the disk--integrated 
degree of circular polarization $P_{\rm c}$ for planets covered by 
those cloud patterns at phase angles ranging from 0$^\circ$ to 180$^\circ$. 
Examples of cloud patterns are given in Fig.~\ref{fig:patches}.

\begin{figure}
\centering
\includegraphics[width=\linewidth]{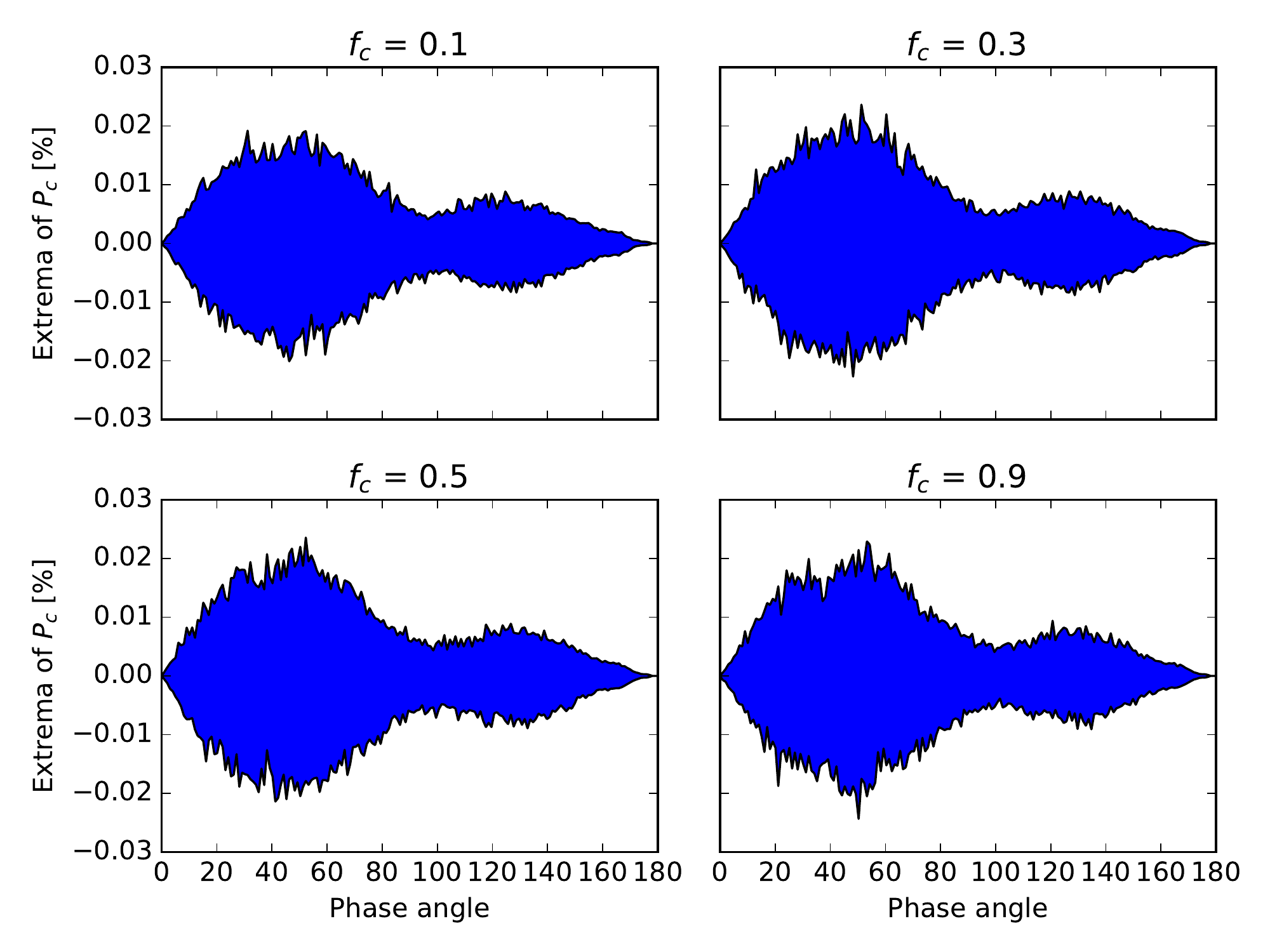}

\includegraphics[width=\linewidth]{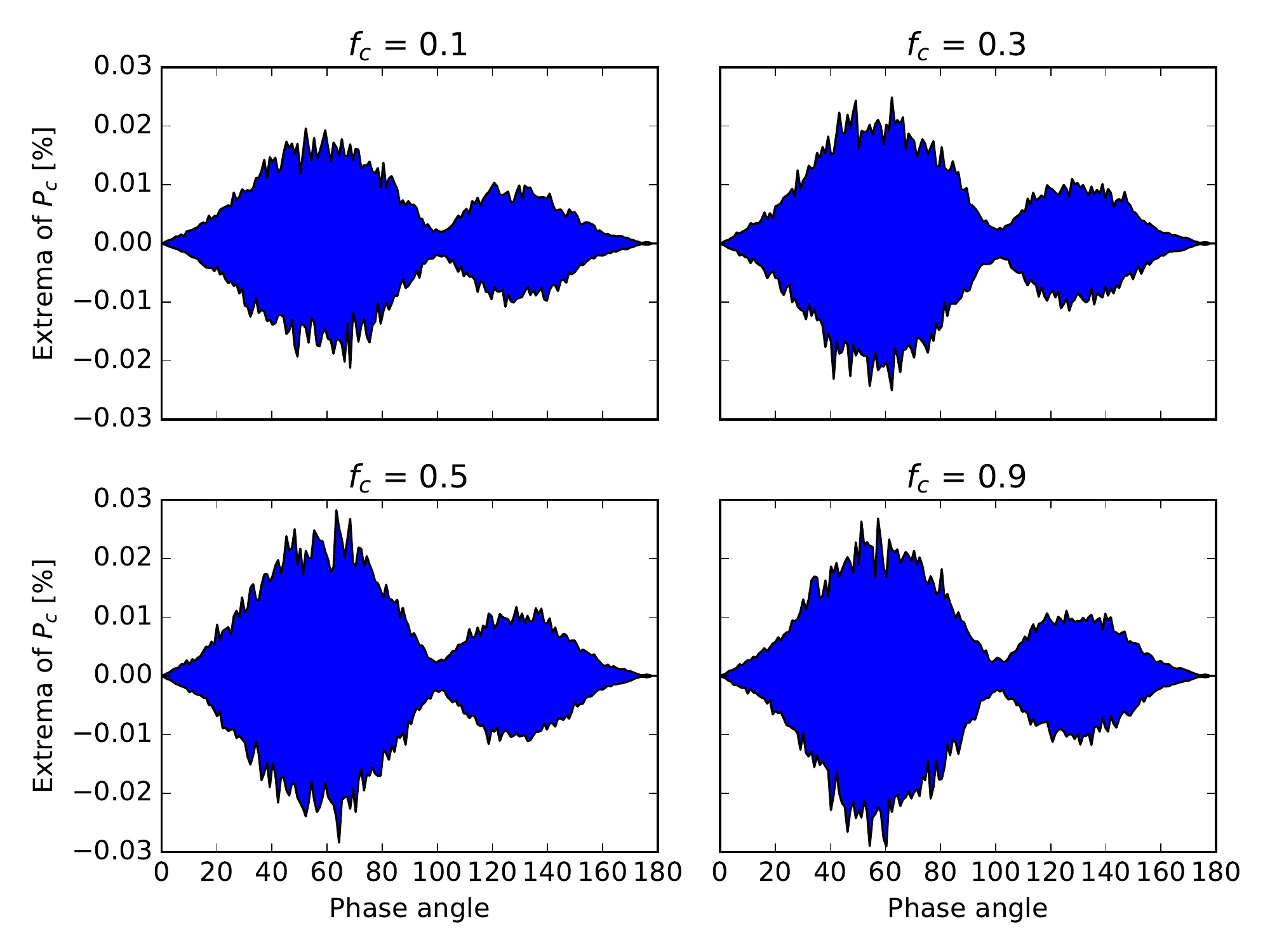}
\caption{Maximum values of $P_{\rm c}$ computed for 300 cloud patterns
         as a function of phase angle $\alpha$, for different cloud
         coverage fractions $f_{\rm c}$, and for Earth--like 
         (top) and Venus--like cloud particles (bottom).}
\label{fig:maxpv-fc}
\end{figure}

Figure~\ref{fig:maxpv-fc} shows as functions of $\alpha$, the maximum and minimum
values of $P_{\rm c}$ as obtained with 300~different cloud patterns for cloud coverage 
fractions $f_{\rm c}$ ranging from 0.1 (10\% clouds) to 0.9 (90\% clouds), 
for Earth--like and Venus--like cloud particles. The first thing to note are 
the overall very small values of $P_{\rm c}$ for these horizontally inhomogeneous 
planets: the maximum values of $P_{\rm c}$ are about 0.020\% for both the Earth--like, 
water clouds particles, and the Venus--like, sulfuric acid cloud 
particles. For both cloud types, these maximum values occur around 
$\alpha= 50^\circ - 60^\circ$. A second, smaller maximum occurs around 
$\alpha= 130^\circ$, also for both cloud types. These (local) extreme values 
of $P_{\rm c}$ appear to be relatively independent of the cloud fraction 
$f_{\rm c}$; only for $f_{\rm c}=0.1$, the maximum values are slightly smaller. 

Measuring $P_{\rm c}$ thus does not seem to provide insight into the
cloud coverage fraction $f_{\rm c}$ of a planet. This might seem surprising,
as on our model planets circularly polarized light is only induced by
scattering by the cloud particles. However, it can be explained by
looking at the occurrence of cloudy pixels on both sides of the planet's
hemisphere, keeping in mind that the circularly polarized flux of a 
cloudy pixel on one hemisphere will be 'cancelled' if the pixel's mirror
pixel on the other hemisphere is also cloudy. 
At small cloud coverage fractions, there are few cloudy pixels to contribute 
circularly polarized flux, while the probability is small that a cloudy pixel
has a cloudy mirror pixel on the opposite hemisphere that would
cancel its polarized flux contribution.
At large cloud coverage fractions, there are many cloudy pixels that contribute 
circularly polarized flux, but the probability that a cloudy pixel
has a cloudy mirror pixel on the opposite hemisphere that cancels its 
polarized flux contribution is also large. 

Apart from the minimum values of $P_{\rm c}$ at $\alpha=0^\circ$ and 
180$^\circ$, which are due to geometrical symmetry 
(note that $P_{\rm c}$ can deviate slightly from 0.0 at these phase angles
due to the horizontal inhomogeneities), $P_{\rm c}$ shows a clear local 
minimum around $\alpha=100^\circ$, with a value of about 0.005\% for the 
Earth--like cloud particles and 0.002\% for the Venus--like cloud particles.
These local minimum values of the disk--integrated $P_{\rm c}$ 
around $\alpha=100^\circ$ are related to the low
spatially resolved values of $P_{\rm c}$ for $\alpha=90^\circ$ in 
Figs.~\ref{fig5} and \ref{fig7}. Phase angles close to 90$^\circ$ are
usually considered to be the best for direct observations of exoplanets
because there the angular distance between a planet and its star 
will be largest.  
Our results show that this phase angle range is, however, far
from optimal for measurements of the circular polarization of exoplanets.
The positive aspect of the low values of $P_{\rm c}$ in this phase angle
range is that the influence of circularly polarized light on measurements
of linear polarization, for example\ through cross--talk between optical elements
in an instrument or telescope, will be minimum in this phase angle range. 
Indeed, this phase angle range is in particular interesting for measurements
of the state of the linear polarization of exoplanets with Rayleigh scattering atmospheres, for which $P_{\rm l}$ will be large 
\citep[][]{Seager2000,Stam2004,Stam2008}.
 
The small differences in the general shape of the disk-integrated $P_{\rm c}$
as a function of $\alpha$ of the planets that are covered with water--clouds
and those that are covered by sulfuric acid clouds, in particularly at small
and large phase angles, are indicative of
the different light scattering properties of the cloud particles
(cf. Fig.~\ref{fig:single}),
but will likely be far too subtle to be useful for cloud particle 
characterization. Indeed, the state of linear polarization, which has been
proven to hold such information \citep[see e.g.][]{Hansen1974a,Hansen1974}, 
is expected to be much larger than $P_{\rm c}$ at most phase angles. The
exception being the phase angles where $P_{\rm l}=0$ (i.e. the so--called neutral points) but $P_{\rm c} \neq 0$.  

\begin{figure}
\centering
\includegraphics[width=\linewidth]{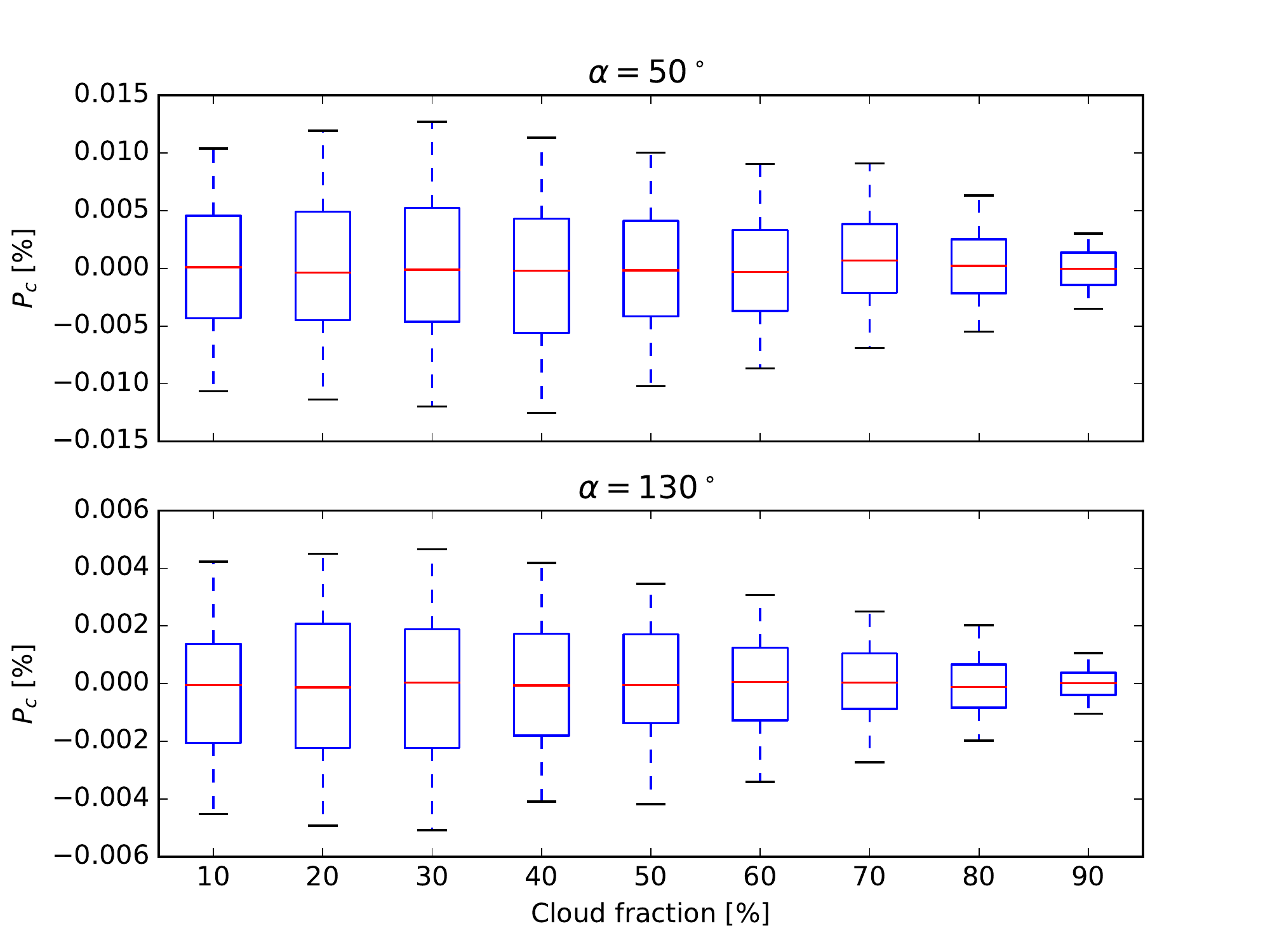}

\includegraphics[width=\linewidth]{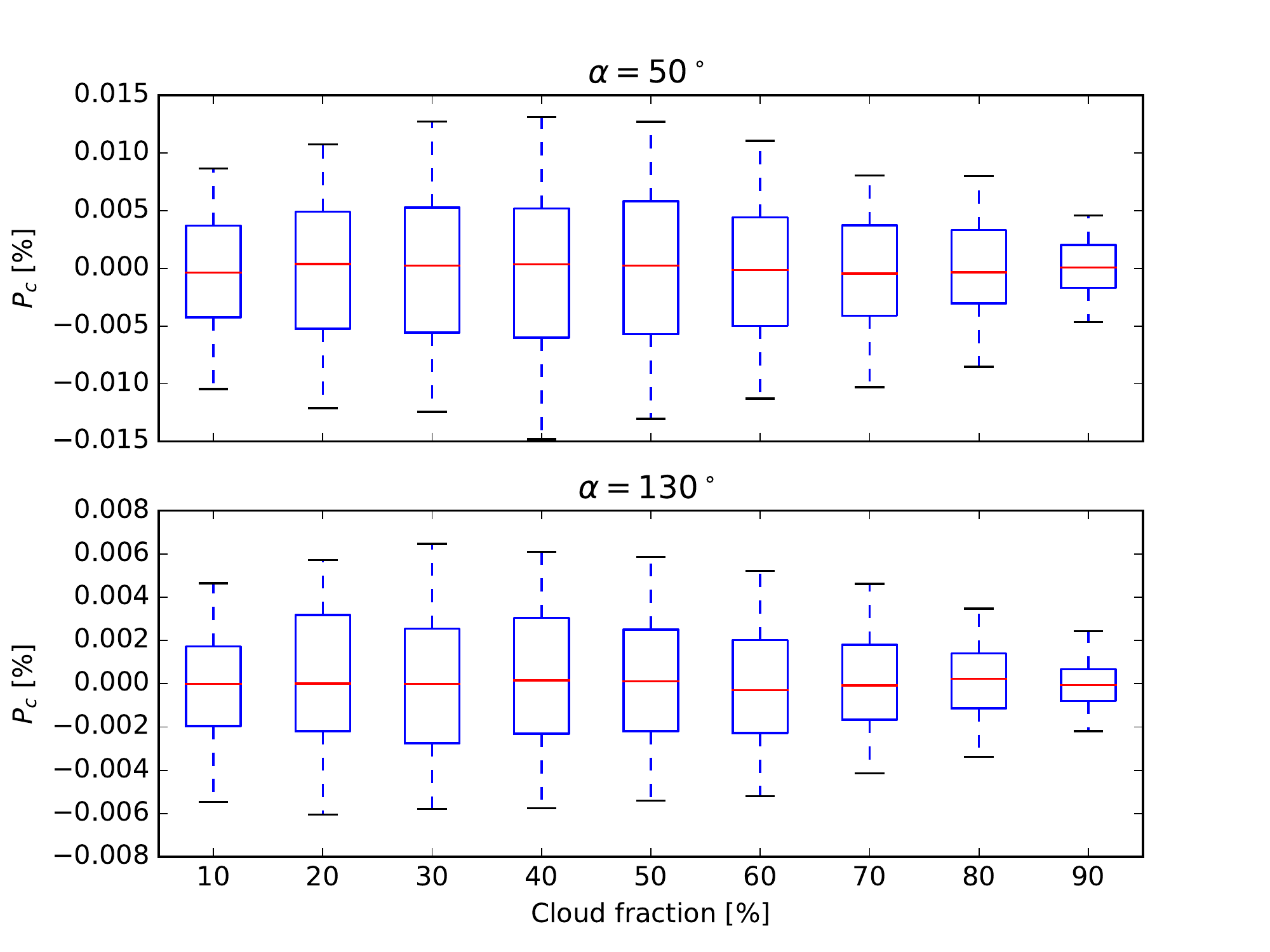}
\caption{Box--plots of the distribution of $P_{\rm c}$ over 300 cloud patterns, 
         for $\alpha=50^\circ$ and $130^\circ$, and the Earth--like (top)
         and Venus--like (bottom) cloud particles. The box indicates the first 
         and third quartiles, with the median in red. The whiskers show
         the fifth and $95^{th}$ percentiles. }
\label{fig:boxplots}
\end{figure}

Figure~\ref{fig:boxplots} shows the statistics of $P_{\rm c}$ 
computed for the 300 cloud patterns as functions of the cloud fraction 
$f_{\rm c}$ for $\alpha=50^\circ$ and 130$^\circ$, the phase angles 
where the maximum values of $P_{\rm c}$ occur in Fig.~\ref{fig:maxpv-fc}. 
It can be seen that the distributions 
are quite symmetrical and have a median quite close to the average value
(that is equal to 0.0\%), as it should be because the clouds can appear
on both hemispheres.
As can be seen, the range of values of $P_{\rm c}$ increases with increasing 
cloud coverage fraction $f_{\rm c}$ from 0.1 to 0.4 - 0.5. Then
it decreases with increasing $f_{\rm c}$, and it is particularly small
for $f_{\rm c}$=0.9, for both phase angles.
The very small range in $P_{\rm c}$ at the largest values of $f_{\rm c}$
is due to the planetary disk being almost completely cloudy and thus 
almost symmetric. At $\alpha=130^\circ$, only a narrow crescent of the
planetary disk is illuminated and visible and the symmetry is even
higher, and the range of $P_{\rm c}$ even smaller, 
than at $\alpha=50^\circ$. 
At $f_{\rm c}=0.1$ (10\%), on the other hand, the disk
is almost devoid of cloudy pixels and the range of $P_{\rm c}$ is thus small.
Like in Fig.~\ref{fig:maxpv-fc}, the dependence of the range 
of $P_{\rm c}$ on $f_{\rm c}$ at intermediate cloud coverages
is small.

\section{Discussion}
\label{sec:discussion}

The results of our simulations show that the degree of circular polarization of
planets with Earth-- or Venus--like cloud particles is really small, especially
compared to the linear polarization. We address here some limitations of this
study and potential issues that observers trying to measure the circular
polarization signals could face.

\subsection{Effects of non-Lambertian reflecting surfaces}

For a solid planet, the albedo $A_{\rm s}$ of the surface has a minor influence
on the pattern of $P_{\rm c}$ across a cloudy planet. Increasing $A_{\rm s}$
generally decreases $P_{\rm c}$ as the surface reflects isotropically
distributed and unpolarized light upwards.  Even upon scattering by the cloud
particles, this light will not add any net circularly polarized flux to the
pattern of $P_{\rm c}$.

A linearly and/or circularly polarizing surface could produce a non--zero
planetary $P_{\rm c}$, but only if the directional pattern of the linearly
and/or circularly polarized flux is azimuthally anisotropic, if the polarizing
surface covers a significant part of the planet, and if the diffuse sky
flux that is incident on the surface is small compared to the incident direct
sunlight or starlight. Due to the lack of azimuthally anisotropically reflecting
polarized surface models, the model planets in this paper have Lambertian,
that\ is depolarizing and isotropically reflecting, surfaces. 

A thorough investigation of the actual strength of this effect requires a
realistic surface reflection model, which is not available now.  Circular
polarization measurements have been proposed as a technique to search for
chiral molecules that on Earth are related to the presence of life
\citep[][]{Bailey2000,2017JQSRT.189..303P,2012P&SS...72..111S}.  Implementing a
realistic surface reflection model, possibly based on circular polarization
observations of light reflected by (regions on) the Earth or in a laboratory
\citep{2017JQSRT.189..303P}, thus appears to be a direction worthy of future
research.

\subsection{Sources of noise}

Because we do not include background stellar flux in our computations of the
degree of circular polarization, $P_{\rm c}$, the exoplanetary signals
presented here apply to planets that are spatially resolved from their star.
Our results can be adapted for background stellar flux at the location of the
planet: if the stellar flux can be considered to be unpolarized it has to be
added to the total planetary flux. It will thus lower the observable $P_{\rm
c}$. 

Circular polarization from the interstellar medium or from the zodiacal light
could also be a possible source of noise when trying to measure signals from an
exoplanet.  The zodiacal light is known to present circular polarization, as
shown by \citet{Wolstencroft1972}, with degrees of circular polarization up to
$-0.8\%$.  We note that the visible--range circular polarization of the zodiacal light varies
strongly with the solar longitude and that such measurements have not been
reproduced since, leaving our understanding of the circular polarization of the
zodiacal light incomplete.  The interstellar medium (ISM) also produces
polarized light, as non--spherical dust grains can be aligned along lines of magnetic fields,
introducing dichroism and birefringence. The amount of circularly polarized
light from the ISM shows values on the order of $10^{-4}$ \citep{Kemp1972,
Avery1975}, which is close to our values for disk--integrated signals.

Nevertheless, the spectral dependence of the polarization due to dust in the
visible is rather well known \citep{Jones2015}. No significant changes are
expected to occur on the timescale of observations of an exoplanet, especially when
considering that the circular polarization of the planet will vary along the
planetary orbit, which could help identify the planetary signal.

\subsection{Gaseous absorption}

This study ignores the effects of gaseous absorption. The
degree of circular polarization in an absorption band could be different from
that at continuum wavelengths, because more absorption implies less
multiple scattering and a signal originating
in higher atmospheric layers. These effects could yield less circular
polarization, first because circular polarization arises from multiple
scattering but also because higher layers are likely to be pure gas, which
does not produce circular polarization. Less multiple scattering would also
yield less unpolarized light. So the total degree of circular polarization 
in an absorption band could also increase with respect to that of the continuum.
The change of $P_{\rm c}$ across an absorption band would thus critically
depend on the amount of absorption and on the structure and composition
of the atmosphere, just like the change of the degree of linear polarization
across absorption bands, as can be seen in for example
\citet{Fauchez2017} and \citet{Stam1999}. A careful investigation of the behaviour of
$P_{\rm c}$ in the presence of gaseous absorption requires a
dedicated study when circular polarization observations
in spectral regions with absorption bands are being planned, 
and is outside the scope of this paper.


\section{Summary}
\label{sec:summary}

We have computed the degree and direction of the circular polarization of planets,
both spatially resolved and disk--integrated.  The former would be applicable
to solar system planets and the latter to exoplanets.
Because sunlight or starlight that is incident on a planet is not circularly
polarized, purely gaseous atmospheres do not yield any circularly polarized
reflected light \citep[cf. the single scattering matrix of Rayleigh scattering
in][]{Hansen1974}, so all of our planets have cloudy atmospheres. Multiple
scattering of incident unpolarized sunlight or starlight by gaseous molecules and
then by cloud particles, or between cloud particles, will usually produce a
circularly polarized signal. 

Spatially resolved simulations of our horizontally 
homogeneous, cloudy model planets show that $P_{\rm c}$ 
is about a factor of 100 smaller than the degree of linear polarization
$P_{\rm l}$. 
Furthermore, $P_{\rm c}$ is mirror symmetric with respect to a planet's light equator 
(which coincides with the equatorial plane on our planets), 
except with the opposite sign. 
On the light equator, $P_{\rm c}$ is zero.

The pattern of $P_{\rm c}$ across the planetary disk depends on the type of
cloud particles, and on the gaseous scattering optical thickness $b_{\rm m}$
above, within, and below the clouds, as gas scatters linearly polarized light
onto the cloud particles that will subsequently be scattered as (partly)
circularly polarized light. For both the water and the sulfuric acid particles
comprising our model clouds, a significant amount of gas above the clouds
yields positive (negative) $P_{\rm c}$ across the northern (southern)
hemisphere at small phase angles, and negative (positive) $P_{\rm c}$ across
the northern (southern) hemisphere at large phase angles. The sign change
appears to happen at phase angles around $90^\circ$, except for very small
cloud optical thicknesses ($b_{\rm c}=0.5$ in Fig.~\ref{fig7}).

Increasing the cloud optical thickness $b_{\rm c}$ increases the amount of
light scattered between the cloud particles, and hence $P_{\rm c}$, for $b_{\rm
c}$ up to about 2.0. Increasing $b_{\rm c}$ even further appears to decrease
$P_{\rm c}$ slightly because then the multiple scattering appears to add mostly
to the total flux $F$.
With decreasing $b_{\rm m}$ above the clouds (i.e.\ with increasing cloud top
altitude and/or increasing wavelength), the pattern of $P_{\rm c}$ due to
the scattering of light between cloud particles becomes dominant. For our model
clouds, this pattern shows sign changes and thus regions of zero $P_{\rm c}$
across each hemisphere. 

For horizontally homogeneous planets, the disk--integrated value of $P_{\rm c}$
is zero, because the circular polarization on the southern hemisphere mirrors
that on the northern hemisphere, except with opposite sign. For a horizontally
inhomogeneous planet, such as a planet covered by patchy clouds, the
disk--integrated $P_{\rm c}$ is usually not equal to zero.  Knowing the values
that can be expected for the disk--integrated $P_{\rm c}$ is important for
exoplanetary detections. We therefore investigated the range of values that the
disk--integrated $P_{\rm c}$ might attain for model planets covered by patchy
clouds of two different compositions, with different cloud coverage percentages
$f_{\rm c}$, and different spatial cloud patterns. 

For various values of $f_{\rm c}$ and 300~different cloud patterns, the maximum
values of the disk--integrated $P_{\rm c}$ are about 0.020\% for both
Earth--like and Venus--like cloud particles.  These maximum values occur at
phase angles $\alpha$ between about 50$^\circ$ -- 60$^\circ$. A secondary,
smaller maximum occurs around 130$^\circ$.  At $\alpha=0^\circ$ and
180$^\circ$, the disk--integrated $P_{\rm c}$ equals zero due to symmetry (for
the horizontally inhomogeneous planet, it can deviate slightly from zero).
Another minimum of $P_{\rm c}$ is smaller than 0.005\% and occurs around
$\alpha=100^\circ$, both for the Earth--like and the Venus--like cloud
particles. The relative independence of these values to $f_{\rm c}$, the cloud
pattern, and the cloud particle type implies that measuring $P_{\rm c}$ of an
exoplanet will provide little information on the cloud properties. 

In the phase angle range where the disk--integrated $P_{\rm c}$ is relatively
small,\ around $\alpha=100^\circ$, $P_{\rm l}$, the degree of linear
polarization, of a planet with an atmosphere in which gaseous molecules are the
main scatterers, will actually be relatively large
\citep[][]{Stam2004,Stam2008}.  In this phase angle range, an exoplanet in a
(more or less) circular orbit will have the largest angular distance from its
star and thus be a good target for direct measurements of the exoplanet's flux
and/or linear polarization signal. Our simulations show that when designing
instruments for such measurements, the influence of the circular polarization
signal, for example\ through cross--talk in the optical components, can likely be
ignored.

With the aim of this study being to investigate what signals can be observed and
their physical origin, we do not explicitly address here the use of circular polarization
for characterization of the atmosphere.  Measuring the patterns in circular
polarization could in principle provide information about the cloud particle
properties (size, composition, shape) and atmospheric structure (cloud
altitude, thickness). Further research is needed to investigate which
information $P_{\rm c}$ would provide that could not be obtained from flux and
in particular linear polarization measurements, as $P_{\rm l}$ is usually much
larger than $P_{\rm c}$, and which accuracy would be required for such $P_{\rm
c}$ measurements in order to be able to derive such additional information.

Spatially resolved observations of the visible and infrared circular
polarization of the Earth from space would help to validate the radiative
transfer codes, our results for the cloudy regions on the planet, and to
provide measurements that can be used to develop a realistic circularly
polarizing surface model to improve the codes. Such measurements could
also be done in a laboratory setting, provided a range of illumination and
viewing angles (also including the azimuthal direction) is covered.

\begin{acknowledgements}
L. Rossi acknowledges funding through the PEPSci Programme of NWO, the Netherlands
Organisation for Scientific Research.
\end{acknowledgements}

\bibliographystyle{aa}
\bibliography{biblio}


\begin{appendix}

\section{Comparison with \citet{Kawata1978}}
\label{app:kawata}

The spatially resolved signals computed in Section \ref{sect:homogeneous} can be compared with those in Figs.~11, 12, and 13
of \citet{Kawata1978}. The comparison is limited since \citet{Kawata1978} used no Earth--like cloud
particles. The phase angle is, however, $53^\circ$, which is close to our value 
of $60^\circ$. 
Kawata's cloud particles have the same effective radius and variance as our
Venus--like cloud particles, but since he used different wavelengths from us,
the refractive index is different (i.e.\ 1.46 at 340~nm, 1.433 at 700~nm,
and 1.43 at 1000~nm).
\citeauthor{Kawata1978} used three atmosphere models. Model~1 has 
one layer of mixed gas and cloud particles. The gas scattering optical thickness at
365~nm is 6, and the cloud optical thickness 128.
Model~2 consists of a lower layer with only cloud particles and an upper layer of gas.
Model~3 has a lower layer with mixed cloud particles and gas, and an 
upper layer of gas.

\citeauthor{Kawata1978} compares the signals of these model atmospheres at 
380~nm where the gas has a large influence, and at 700~nm, where the gas has 
little influence.
His results for his Model~1 at 700~nm \citep[Fig.\ 11 in][]{Kawata1978} should
be comparable to those of our atmosphere Models~B (the gas optical thickness
in our atmospheric layers is only 0.036) and~D with Venus--like particles.
Indeed, our results show similar characteristics to those of \citeauthor{Kawata1978}:
in the absence of significant scattering by gas above the clouds, 
the neutral band crossing the disk from north to south is similar (although in 
our model, the band is shifted somewhat to the limb), as are
the signs of $P_{\rm c}$. We also find the increase of $P_{\rm c}$ towards to limb.
Our Model~D shows an additional neutral band close to the terminator that is
absent in the results of \citeauthor{Kawata1978}. This difference is probably due
to the slight differences in atmosphere models, refractive index, depolarization
factor for the Rayleigh scattering, and/or phase angle.

The results of \citeauthor{Kawata1978} in case the contribution of gas above the clouds
is significant (his Fig.\ 12) are also similar to ours, except that in our model atmospheres, 
the gas optical thickness is much smaller: the presence of gas removes the north--south
neutral band, leaving mostly positive $P_{\rm c}$ across the northern hemisphere,
with the highest values towards the limb. We do not find the slightly negative region 
north of the equator and towards the terminator on the northern hemisphere
(and reversed on the southern hemisphere) that shows up in Kawata's Fig.~12.
Indeed, we adapted our model atmosphere, the cloud particles' properties, and the phase 
angle to simulate 
Kawata's Fig.~12 as closely as possible, and while then the overall shape of $P_{\rm c}$ across 
our planetary disk is very similar to that in Fig.~12 of Kawata, our maximum values 
(at the limb) are only about 0.017\% while those found by Kawata appear to be higher 
than 0.020\%. We do find a slightly negatively polarized region on the northern 
hemisphere (and a slightly positively polarized region on the southern hemisphere)
at approximately the same location as in Fig.~12 of Kawata, except with values
that are a factor of ten smaller than those of Kawata. We also find small regions
with inverse polarization near the poles that are absent in Fig.~12 of
\citeauthor{Kawata1978}. The origin of these discrepancies is unknown,
as not all details on Kawata's numerical computations are known to us,
but they are extremely small, in particular in absolute sense, and leave us
confident in our results.

\begin{figure}[b!]
\begin{center}
\includegraphics[width=5.5cm]{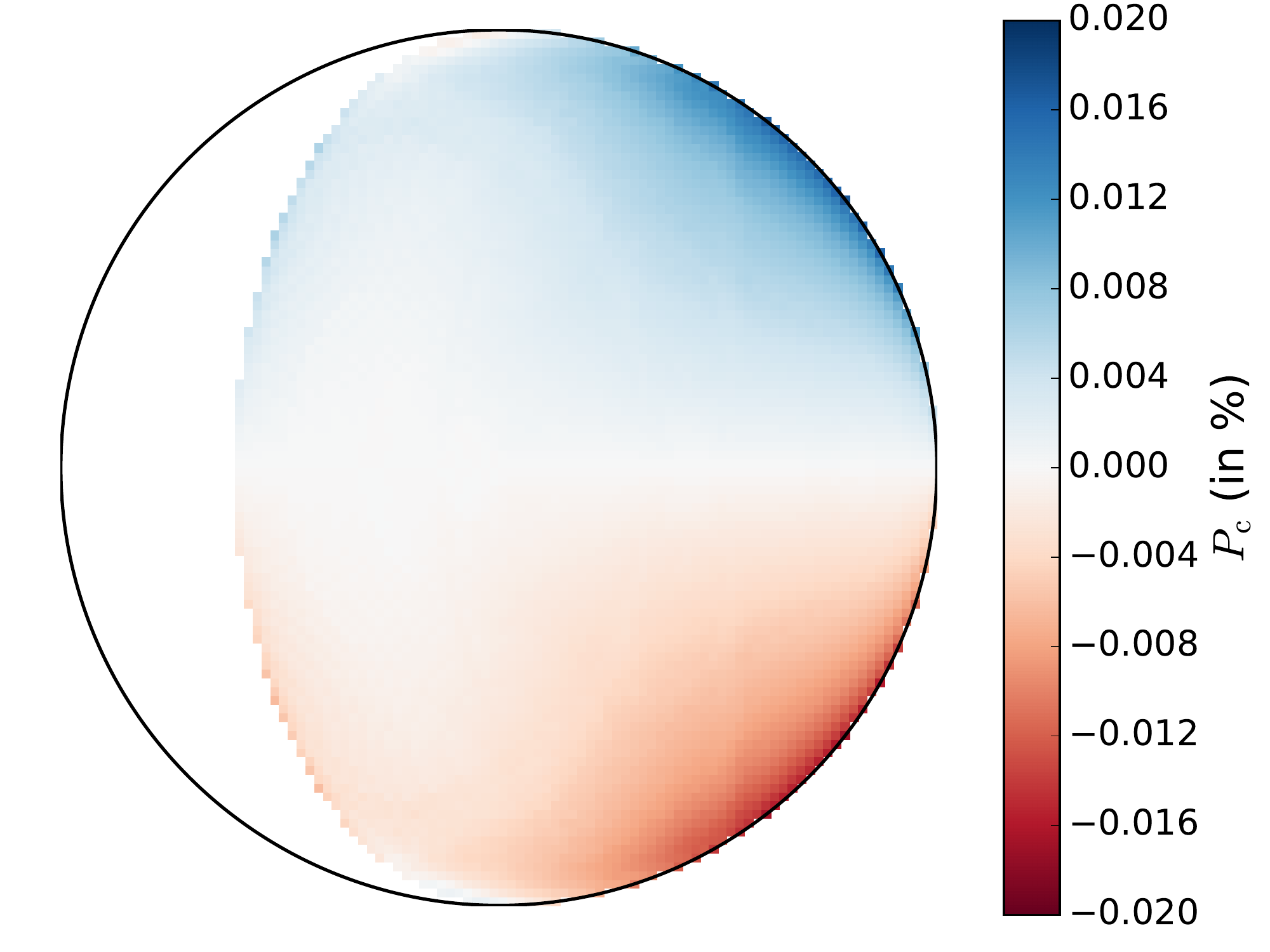}
\end{center}
\caption{Degree of circular polarization $P_{\rm c}$ for $\alpha=53^\circ$
         and a model atmosphere as close as possible to that used in Fig.~12
         by \citeauthor{Kawata1978}: the atmosphere consists of a single
         layer with $b_{\rm m}= 0.51$ and cloud particles with 
         $b_{\rm a}= 128$. The cloud particles are described by a two-parameter
         gamma size distribution with $r_{\rm eff}= 1.05$~$\mu$m and 
         $v_{\rm eff}= 0.07$, with a refractive index equal to 1.458.}
\label{fig_kawata}
\end{figure}

\section{Additional figures}

\begin{figure*}[h]
\begin{center}
\includegraphics[width=16cm]{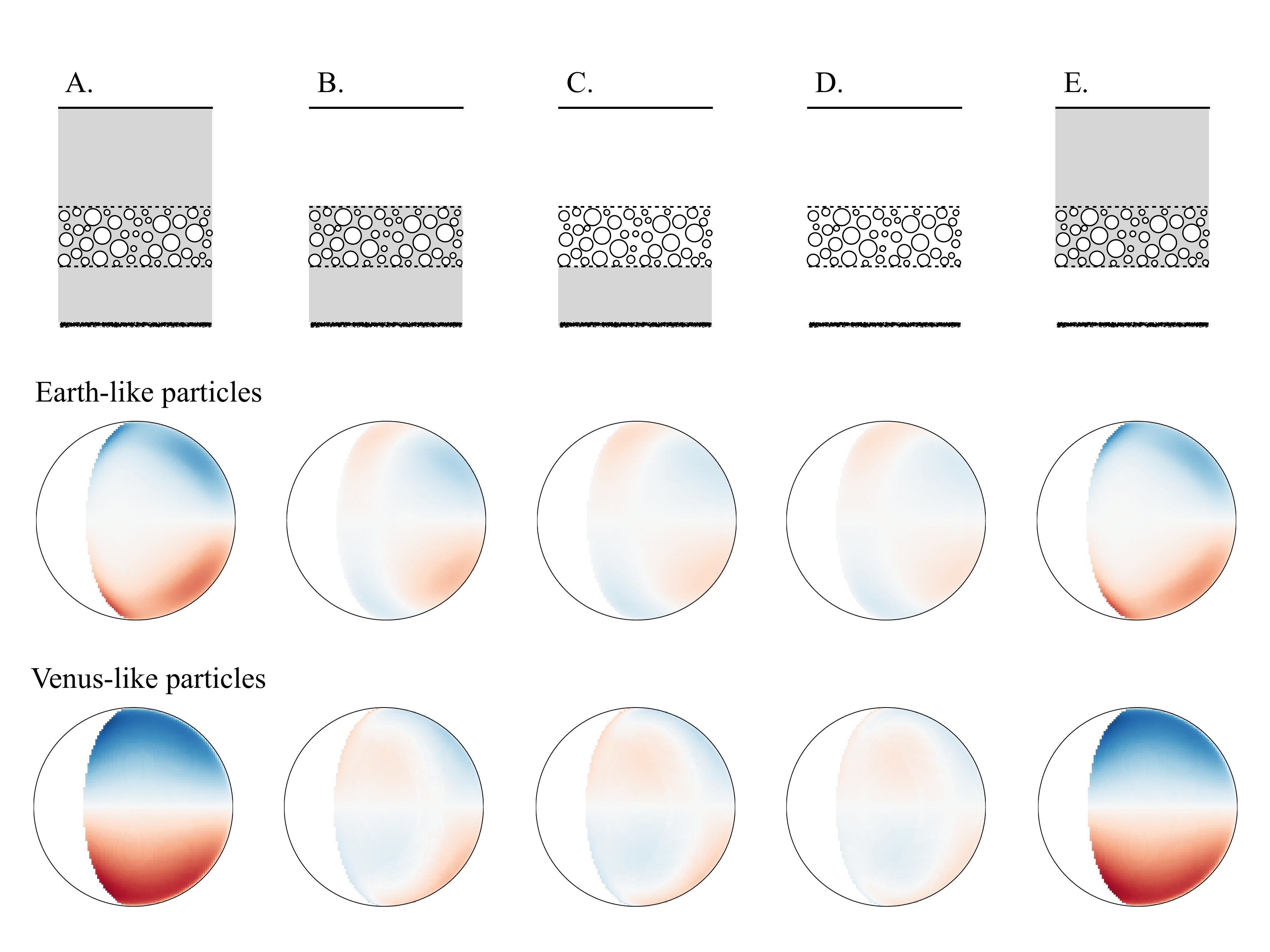}
\end{center}
\caption{Similar to Fig.~\ref{fig3}, except only for $P_{\rm c}$ and  
         five~different atmosphere models. The cloud top is at 4~km, the subdivision
         of the atmosphere above the cloud as in Fig.~\ref{fig:atmos}
         is not shown. 
         From left to right: 
         A. gas in all atmospheric layers (cf.\ Fig.~\ref{fig3}); 
         B. only gas in the bottom and in the cloud layer;
         C. only gas in the bottom layer;
         D. no gas at all;
         E. no gas in the bottom layer.}
\label{fig4}
\end{figure*}

\begin{figure*}[h!]
\begin{center}
\vspace*{-0.5cm}
\includegraphics[width=0.85\linewidth]{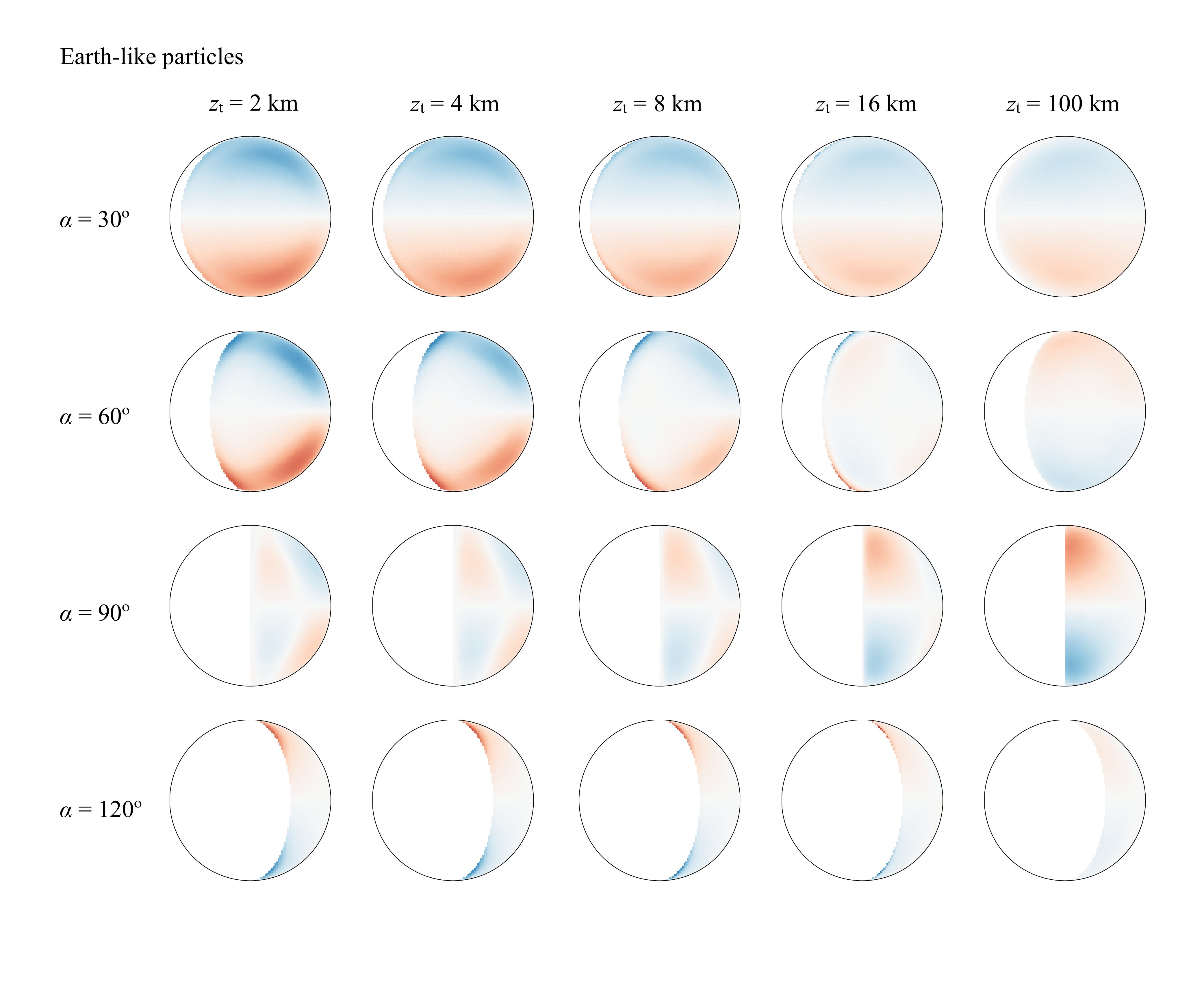} 

\vspace*{-1.5cm}
\includegraphics[width=0.85\linewidth]{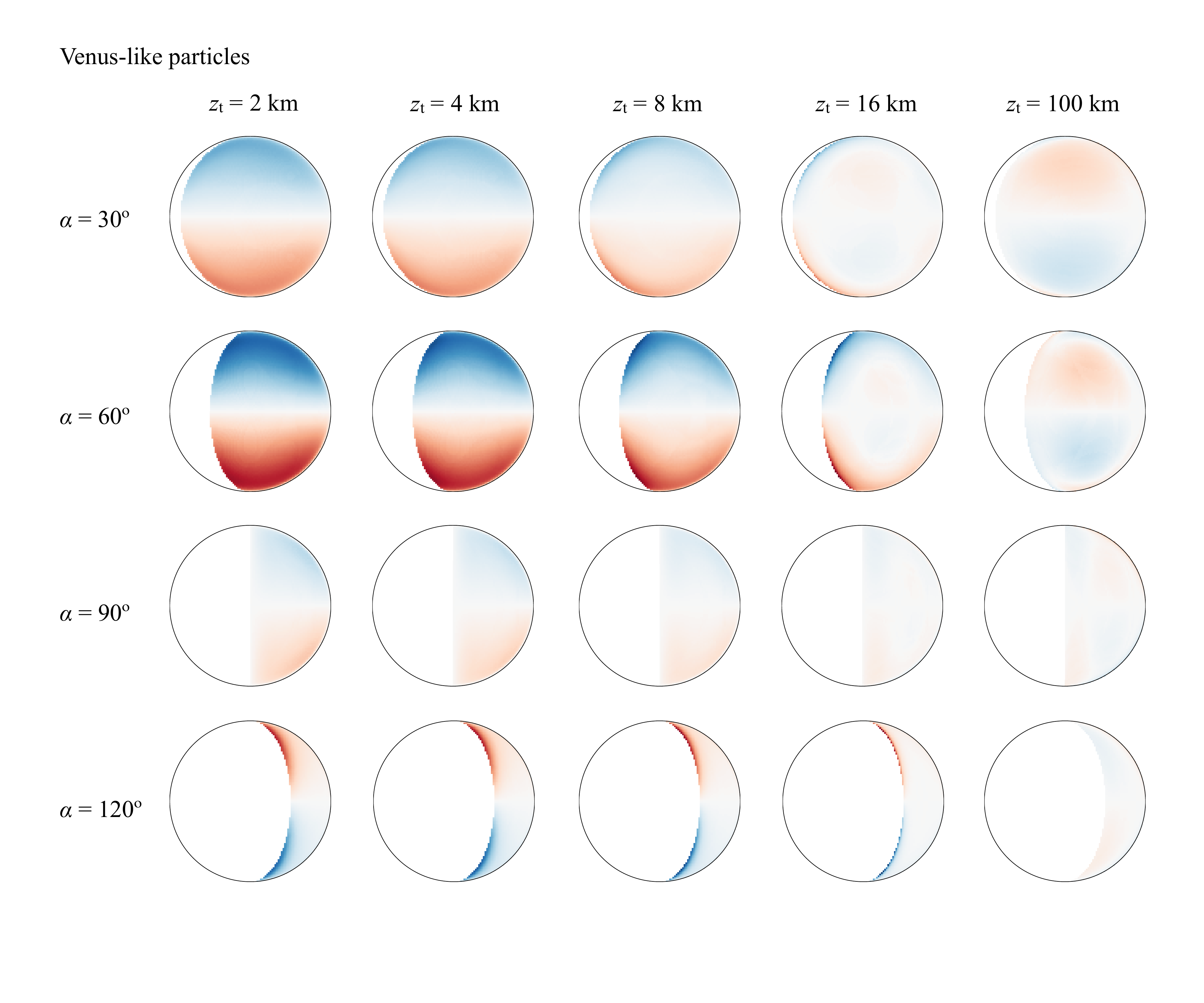} 
\vspace*{-1.5cm}
\end{center}
\caption{Degree of circular polarization $P_{\rm c}$ for $\alpha$ equal to 
         $30^\circ$, $60^\circ$, $90^\circ$, and 120$^\circ$. 
         The standard model atmosphere has a $b_{\rm c}=2.0$
         cloud consisting of Earth--like (top) or Venus--like (bottom) particles.
         The cloud top altitudes $z_{\rm t}$ are:
         2~km (left columns), 4~km (cf.\ Fig.~\ref{fig3}),
         8~km, 16~km, and 100~km (right columns).
         The colour scale is the same as that in Fig.~\ref{fig3}.}
\label{fig5}
\end{figure*}

\begin{figure*}[h]
\begin{center}
\vspace*{-0.5cm}
\includegraphics[width=0.85\linewidth]{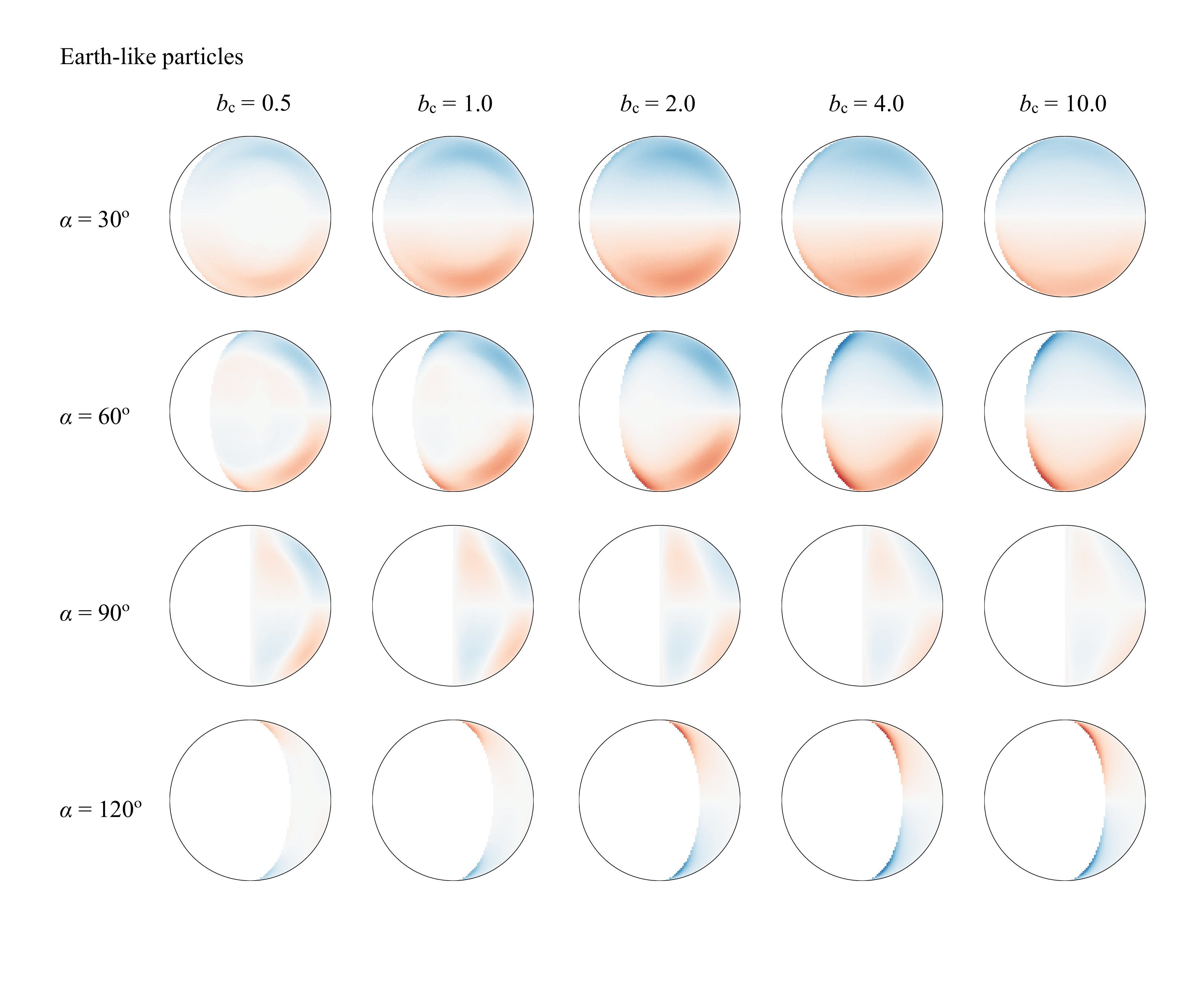} 

\vspace*{-1.5cm}
\includegraphics[width=0.85\linewidth]{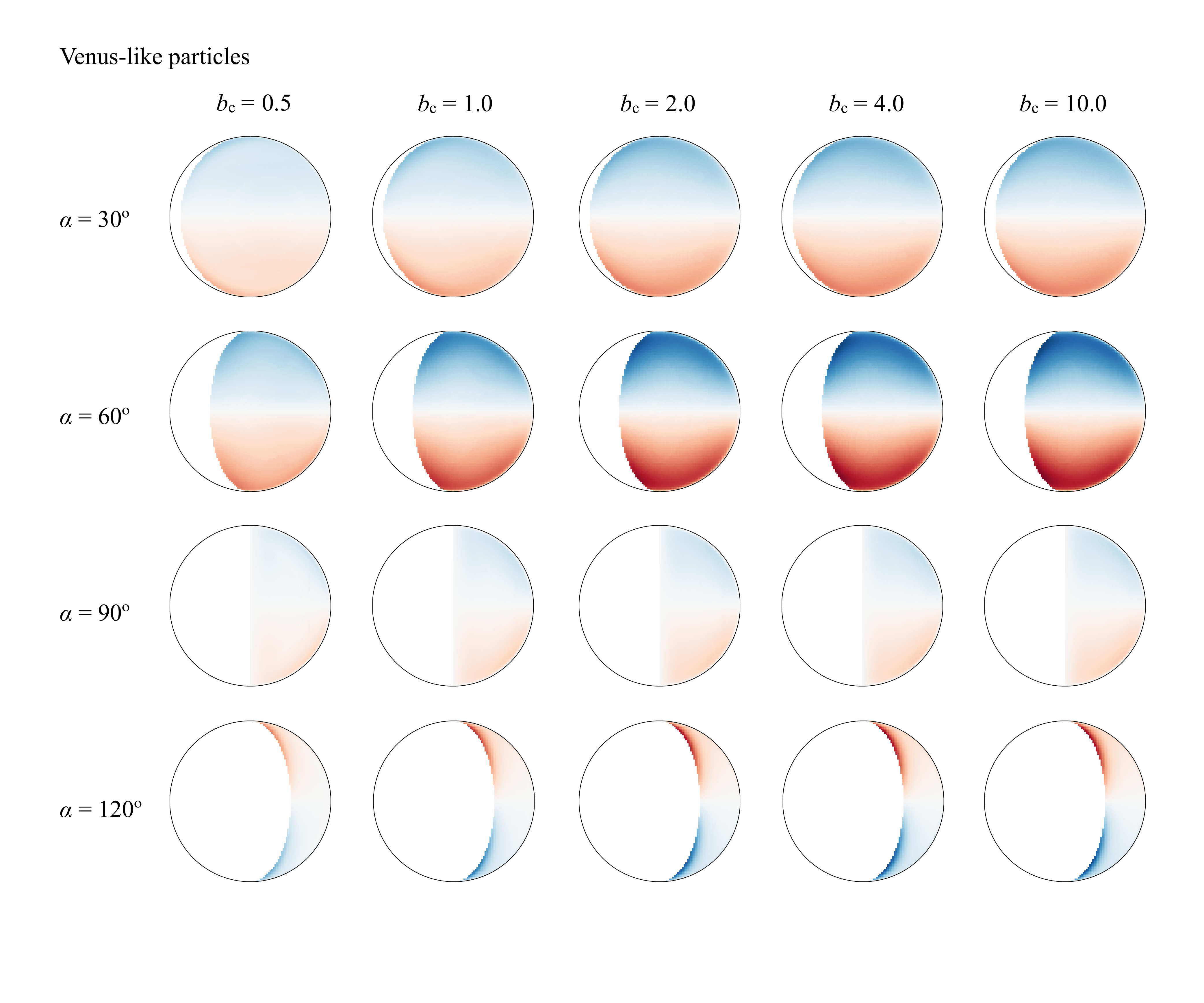} 
\vspace*{-1.5cm}
\end{center}
\caption{Degree of circular polarization $P_{\rm c}$ for $\alpha$ equal to
         $30^\circ$, $60^\circ$, $90^\circ$, and 120$^\circ$.
         The standard model atmosphere has a cloud consisting of 
         Earth--like (top) or Venus--like (bottom) particles,
         with the cloud top altitude $z_{\rm t}=4.0$~km,
         and the following cloud optical thicknesses $b_{\rm c}$:
         0.5 (left columns), 1.0, 2.0, 4.0, and 10.0 (right columns).
         The colour scale is the same as that in Fig.~\ref{fig3}.}
\label{fig7}
\end{figure*}

\end{appendix}

\end{document}